\documentclass[letterpaper,titlepage,11pt]{article}

\usepackage{amssymb,amsmath,amsfonts}
\usepackage{graphicx}
\usepackage{epsfig}
\usepackage{ccaption}
\usepackage{wrapfig}
\usepackage[small,bf,up]{caption}

\usepackage[
      colorlinks=true,
      linkcolor=black,
      urlcolor=black,
      filecolor=black,
      citecolor=black,
      pdfstartview=FitV,
      pdftitle={},
        pdfauthor={Kentaro Tanabe},
        pdfsubject={},
        pdfkeywords={},
        pdfpagemode=None,
        bookmarksopen=true
      ]{hyperref}

\def\clock{{\count0=\time
           \divide\count0 60
           \ifnum\count0<10 0\fi\the\count0
           \multiply\count0 -60 \advance\count0 \time
           :\ifnum\count0<10 0\fi \the\count0
         }}
\newcommand{\timestamp}{{\small\vbox{\hbox{\tt\jobname.tex}
\hbox{\the\day/\the\month/\the\year, \clock}}}}

\newcommand{\sR}{\mathsf{R}}

\newcommand{\tH}{\text{H}}

\newcommand{\hL}{\hat{\Lambda}}

\newcommand{\ca}{\cosh{\alpha}}
\newcommand{\sa}{\sinh{\alpha}}
\newcommand{\ta}{\tanh{\alpha}}
\newcommand{\cca}{\cosh^{2}{\alpha}}
\newcommand{\ssa}{\sinh^{2}{\alpha}}

\newcommand{\sssa}{\sinh^{3}{\alpha}}

\newcommand{\pz}{\partial_{z}}
\newcommand{\pv}{\hat{\partial}_{t}}
\newcommand{\pp}{\partial_{\psi}}

\newcommand{\pt}{\partial_{\theta}}

\newcommand{\bA}{\boldsymbol{A}_{(N)}}
\newcommand{\bJ}{\boldsymbol{J}_{(N)}}
\newcommand{\bPsi}{\boldsymbol{\Psi}}

\numberwithin{equation}{section}

\begin{document}

\begin{titlepage}
\rightline{RUP-16-16} 
\leftline{}
\vskip 2cm
\centerline{\LARGE \bf Charged rotating black holes at large $D$}
\vskip 1.6cm
\centerline{\bf Kentaro Tanabe}
\vskip 0.5cm
\centerline{\sl Department of Physics, Rikkyo University,}
\centerline{\sl  Toshima, Tokyo 171-8501, Japan}
\smallskip
\vskip 0.5cm
\centerline{\small\tt ktanabe@rikkyo.ac.jp}

\vskip 1.6cm
\centerline{\bf Abstract} \vskip 0.2cm \noindent
We study odd dimensional charged equally rotating black holes in the Einstein-Maxwell theory with/without a cosmological constant by using the large $D$ expansion method, where $D$ is a spacetime dimension.  Solving the Einstein-Maxwell equations in the $1/D$ expansion we obtain the large $D$ effective equations for charged equally rotating black holes. The effective equations describe the nonlinear dynamics of charged equally rotating black holes. Especially the perturbation analysis of the effective equations gives analytic formula for quasinormal mode frequencies, and we can show charged equally rotating black holes have instabilities. As one interesting feature of instabilities, we observe that the ultraspinning instability of neutral equally rotating black holes in de Sitter is connected with the instability of de Sitter Reissner-Nordstrom black hole in a rotation-charge plane of the solution parameter space. So these instabilities have same origin as dynamical properties of charged rotating black holes. We also give perturbation analysis by a small charge for even dimensional equally rotating black holes.

\end{titlepage}
\pagestyle{empty}
\small
\tableofcontents
\normalsize
\pagestyle{plain}
\setcounter{page}{1}

\section{Introduction}

We have exact solutions of rotating black holes with/without  a cosmological constant in arbitrary dimensions \cite{Kerr:1963ud,Myers:1986un,Gibbons:2004js,Gibbons:2004uw}. So it may be natural to expect that charged versions of exact solutions in the Einstein-Maxwell theory can be obtained also in a simple analytic expression. This expectation, however, has not been realized yet except for four dimensional solutions, that is, the Kerr-Newman solution \cite{Newman:1965my}. We do not have analytic expressions of higher dimensional version of the Kerr-Newman solution\footnote{If one changes the theory, also the situation changes. For example, allowing the dilaton coupling or Chern-Simons term, we can construct some exact solutions of charged rotating black hole solutions in higher dimensions \cite{Breckenridge:1996is,Breckenridge:1996sn,Llatas:1996gh,Kunz:2006jd}.}. The Einstein-Maxwell equations are nonlinear coupled partial differential equations. Hence it is difficult to find analytic solutions in general, and one needs some nontrivial technique or ansatz to simplify them. In fact the Kerr-Newman solution was found by applying nontrivial relation through the complex transformation between the Schwarzschild and Kerr solution to the Reissner-Nordstrom solution (see \cite{Adamo:2014baa} for a recent review). The extension of this complex transformation to higher dimensions is not straightforward, and there is no guarantee that the transformation exists also in higher dimensions. Another remarkable property of the Kerr-Newman solution, which might play a key role in finding higher dimensional version of the Kerr-Newman solution, is the Kerr-Schild form \cite{Kerr:1965,Debney:1969zz}. In the Kerr-Schild form the gravitational potential of a black hole is linearized in the metric, and various analysis simplifies. The higher dimensional rotating black holes, Myers-Perry black holes, were given by using this Kerr-Schild form \cite{Myers:1986un} . In their original paper \cite{Myers:1986un} they also tried to find higher dimensional charged rotating black holes in the Einstein-Maxwell theory by using the Kerr-Schild form, but failed. After that work, there have been further efforts to find charged rotating black holes in higher dimensions by using the Kerr-Schild form \cite{Fan:2015kte}, perturbation analysis by small charge \cite{NavarroLerida:2007ez,Krtous:2007xg,Aliev:2007qi,Allahverdizadeh:2010xx} or small rotations \cite{Aliev:2005npa,Aliev:2006yk} and other approximation method such as the blackfold method \cite{Caldarelli:2010xz}. There are also numerical work which constructed charged rotating black hole solutions \cite{Kunz:2005nm,Kunz:2006eh,Brihaye:2007bi,Brihaye:2008br,Kunz:2007jq}. In these analysis we can study some physical properties of stationary charged rotating black holes in higher dimensions such as thermodynamic quantities and giro magnetic ratios.    

In this paper we would like to give another analysis on higher dimensional charged rotating black holes in the Einstein-Maxwell theory by the large $D$ expansion method \cite{Emparan:2013moa}. $D$ is a spacetime dimension. In this analysis we can study some physical properties of stationary charged rotating black holes such as thermodynamic quantities. The main target of this paper, however, is  the dynamical property of charged rotating solutions such as instabilities. Higher dimensional black holes sometimes become unstable dynamically. The rotating black holes have instabilities at large rotation so called ultraspinning instabilities \cite{Emparan:2003sy,Dias:2009iu,Dias:2010eu,Dias:2010maa,Dias:2010gk,Dias:2011jg} or bar mode instabilities \cite{Emparan:2003sy,Shibata:2009ad,Shibata:2010wz,Hartnett:2013fba}, and the Reissner-Nordstrom black holes in de Sitter also have the instability at large charge in higher dimensions \cite{Konoplya:2008au,Cardoso:2010rz,Konoplya:2013sba}. We will investigate the charged or rotating version of these instabilities by using the large $D$ expansion method. 

At large $D$ the radial gradients of gravitational potentials become very large with $D$, and, as a result, the gravitational interaction by a black hole is confined in very near region of the black hole horizon. Then the horizon can be effectively regarded as a membrane in a background geometry \cite{Emparan:2015hwa,Bhattacharyya:2015dva,Suzuki:2015iha} for decoupled mode excitations \cite{Emparan:2014aba}. The equations of motion for the membrane can be obtained by integrating the Einstein equations in the radial direction. The obtained effective equations are written by the embedding functions and dynamical quantities of the membrane. The effective equations describe the nonlinear dynamical evolutions of solutions and are very suitable to the stability analysis of black holes. The analysis of the large $D$ expansion method for charged black holes was initiated in \cite{Bhattacharyya:2015fdk} and independently in \cite{Tanabe:2015isb}. In these analysis the decoupled quasinormal modes (light quasinormal modes) of (Anti de Sitter) Reissner-Nordstrom black holes were derived. The dynamics of charged black branes was also studied by the large $D$ expansion method in \cite{Emparan:2016sjk}. The purpose of this paper is to extend our previous analysis \cite{Tanabe:2015isb,Emparan:2016sjk} to charged equally rotating black holes in odd dimensions. Hence we will derive the large $D$ effective equations for charged rotating black holes and perform the stability analysis of solutions. Then we will see that the ultraspinning and bar mode instabilities of rotating black holes in de Sitter are smoothly connected with the instability of de Sitter Reissner-Nordstrom black hole in the parameter space of solutions. Thus these instabilities have common dynamical origin. Note that our analysis is not covered by the formalism developed in \cite{Bhattacharyya:2015fdk}. In our setting we consider black holes with infinite number of non-zero angular momentum at the large $D$ limit, while the formalism in \cite{Bhattacharyya:2015fdk} treats black holes with finite number of angular momentum. 


The organization of this paper is as follows. In section \ref{2} we give the large $D$ effective equations for charged equally rotating black holes. The perturbation analysis of the effective equations are performed in section \ref{3}. This analysis yields the formula for quasinormal mode frequencies of charged equally rotating black holes. We close this paper by giving summary and discussion in section \ref{4}. Some technical details are contained in Appendices \ref{A}. The large $D$ analysis for even dimensional equally rotating black holes with a small charge is given in Appendix \ref{B}. This analysis may be helpful to understand why the large $D$ analysis is simplified for charged equally rotating black holes in odd dimensions.  

We consider $D=2N+3=n+3$ dimensional black hole solutions in the Einstein-Maxwell theory by using the large $D$ expansion method. In this expansion we use $1/n$ as an expansion parameter instead of $1/D$.

\section{Large $D$ effective equations}
\label{2}

In this section we consider the large $D$ effective theory for charged rotating black holes. Especially we concentrate on large $D$ equally rotating solutions in odd dimensions\footnote{The precise definition of large $D$ equally rotating solutions will be given below.}. This is because the exact solution of neutral equally rotating black holes in odd dimensions has a very simple expression due to its cohomogeneity-1 property and we expect that the charged solution has similar property. As for even dimensional solutions, we give some perturbation analysis by a small charge in Appendix \ref{B}. 

\subsection{Large $D$ solutions}

We solve the Einstein-Maxwell equations with a cosmological constant
%
\begin{eqnarray}
R_{\mu\nu}-\frac{1}{2}Rg_{\mu\nu}+\Lambda g_{\mu\nu} = \frac{1}{2}\left( 
F_{\mu\rho}F^{\rho}{}_{\nu}-\frac{1}{4}F_{\rho\sigma}F^{\rho\sigma}g_{\mu\nu}
\right),~~\nabla^{\mu}F_{\mu\nu}=0,
\label{EMeq}
\end{eqnarray}
%
in $D=2N+3$ dimensions. Here $F_{\mu\nu}=\partial_{\mu}A_{\nu}-\partial_{\nu}A_{\mu}$. We consider large $D$ equally rotating solutions in odd dimensions. At first it is instructive to observe the exact neutral solution, which is the equally rotating Myers-Perry black hole in (Anti) de Sitter, whose metric is given in Eddington-Finkelstein coordinates\footnote{Our radial coordinate $r$ is related with the radial coordinate $r_{\text{BL}}$ in the Boyer-Lindquist coordinate \cite{Myers:1986un,Gibbons:2004js,Gibbons:2004uw}  by $r^{2}=(r^{2}_{\text{BL}}+a^{2})/(1+a^{2}\hL)$.} by
%
\begin{eqnarray}
ds^{2} &=& \frac{2r}{\sqrt{r^{2}-a^{2}+a^{2}r^{2}\hat{\Lambda}}}\left( dt-a \bPsi \right)dr 
-\left(1-r^{2}\hat{\Lambda}-\left( \frac{r_{0}}{r} \right)^{n}\right)dt^{2}  \notag \\
&&~~~~~~~~~~~~~~~~~~~~
-a\left(\frac{r_{0}}{r}\right)^{n}\bPsi dt +\left( r^{2}+a^{2}\left(\frac{r_{0}}{r}\right)^{n} \right)\bPsi^{2}+r^{2}d\Sigma^{2}_{N}
\label{MPsol}
\end{eqnarray}
%
where $n=2N$. We introduce the reduced cosmological constant $\hat{\Lambda}$ by
%
\begin{eqnarray}
\Lambda =\frac{(n+1)(n+2)}{2}\hat{\Lambda}.
\end{eqnarray}
%
$d\Sigma_{N}^{2}$ is the Fubini-Study metric on $\mathbb{CP}^{N}$, and $\bPsi$ is a $1$-form defined by
%
\begin{eqnarray}
\bPsi = d\psi + \bA.
\end{eqnarray}
%
$\psi$ is one of rotating directions, and $\partial_{\psi}$ is a Killing vector. The 1-form $\bA$ gives the Kahler form $\bJ$ on $\mathbb{CP}^{N}$ by $2\bJ=d\bA$. $a$ is a rotation parameter, and $r_{0}$ is a horizon radius of the solution. One important observation on this exact solution is the boost property, which appears at large $D$ \cite{Emparan:2014jca}\footnote{ In the paper \cite{Emparan:2014jca} the curvature scale $L=|\hat{\Lambda}^{-1/2}|$ was used for a cosmological constant. }. Introducing a new radial coordinate $\sR$ by
%
\begin{eqnarray}
\sR=\left( \frac{r}{r_{0}} \right)^{n}, 
\label{Rdef}
\end{eqnarray}
%
and we take the large $D$ limit with fixed $\sR$ and $\hat{\Lambda}$. Then the metric (\ref{MPsol}) reduces to the following simpler form at the leading order of the large $D$ limit
%
\begin{eqnarray}
ds^{2} = \eta_{AB}dx^{A}dx^{B} + \frac{\sR_{0}}{\sR}u_{A}u_{B}dx^{A}dx^{B} +2V^{-1}u_{A}dx^{A}dr + d\Sigma^{2}_{N},
\label{MPsolLD}
\end{eqnarray}
%
where we set to $r_{0}=1+O(1/n)$. $\eta_{AB}$ is a two dimensional flat metric with Lorentzian signature. Here we defined $dx^{A}=(Vdt,\bPsi)$, and $V$ satisfies
%
\begin{eqnarray}
V^{2} =1-\hat{\Lambda}.
\label{Vdef}
\end{eqnarray}
%
Note that $V$ vanishes at $\hat{\Lambda}=1$. This limit corresponds to the Nariai limit, where the cosmological horizon and event horizon coincide. For a positive cosmological constant case $\hat{\Lambda}>0$, we always consider small black holes $\hat{\Lambda} \leq 1$. The vector field $u_{A}$ is given by
%
\begin{eqnarray}
u_{A}dx^{A} = V \cosh{\alpha}\,dt-\sinh{\alpha}\,\bPsi.
\end{eqnarray}
%
$\alpha$ is a constant boost parameter of the solution given by
%
\begin{eqnarray}
\tanh{\alpha}=a\sqrt{1-\hat{\Lambda}}.
\label{alphadef}
\end{eqnarray}
%
$\sR_{0}$ is given by
%
\begin{eqnarray}
\sR_{0} = \frac{1-a^{2}+a^{2}\hat{\Lambda}}{1-\hat{\Lambda}}=V^{-2}\cosh^{-2}{\alpha}.
\label{R0def}
\end{eqnarray}
%
$\sR_{0}$ can be absorbed into $O(1/n)$ redefinition of $r_{0}$, so we can set $\sR_{0}=1$ without any loss of generality. Then the effects of rotations in the leading order metric is all represented by the constant boost velocity $\alpha$. If we set to $\alpha=0$, the solution becomes the Schwarzschild black hole in (Anti) de Sitter. Thus the leading order metric of equally rotating Myers-Perry black hole (\ref{MPsol}) in (Anti) de Sitter at large $D$ can be generated by the homogeneous boost transformation on $\mathbb{CP}^{N}$ from the corresponding Schwarzschild black hole. The boost transformation is given by
%
\begin{eqnarray}
dt \rightarrow d\bar{t}=\ca \, dt -V^{-1}\sa\,\bPsi,~~
\bPsi \rightarrow \bar{\bPsi}=\ca \, \bPsi - V\sa\,dt.
\end{eqnarray}
%
$(Vd\bar{t},\bar{\bPsi})$ is the rest frame of the solution. This boost property also exists in singly rotating Myers-Perry black hole \cite{Suzuki:2015iha,Emparan:2013xia}, and the boost is inhomogeneous on the horizon in that case. The appearance of the boost property implies that the various properties of the Schwarzschild black hole are shared with the Myers-Perry black hole at large $D$. In the papers \cite{Suzuki:2015iha, Emparan:2014jca} decoupled quasinormal modes of singly rotating Myers-Perry black holes were obtained analytically by using this property\footnote{
The decoupled quasinormal mode frequency does not obey the boost transformation since the frequencies are obtained from the subleading corrections in the $1/D$ expansion which does not posses the boost property. However the decoupling property of the perturbation variables of the Schwarzschild black hole holds also for Myers-Perry black holes not only at leading order but also for subleading corrections. Then we can solve perturbations equations by similar ways with for Schwarzschild black holes, and we can find quasinormal mode frequencies analytically. 
}. Here we employ this boost property of rotating large $D$ black holes to construct the charged equally rotating black holes. 

We consider dynamical deformations of the exact solution (\ref{MPsol}) which break $\mathbb{CP}^{N}$ symmetry. To this end we use the fact that $d\Sigma_{N}^{2}$  is decomposed as
%
\begin{eqnarray}
d\Sigma^{2}_{N} = d\theta^{2} + \cot^{2}{\theta}\,\bA^{2} +\sin^{2}{\theta}\,d\Sigma^{2}_{N-1}.
\label{CPN}
\end{eqnarray}
%
$\theta$ is one of non-Killing coordinates on $\mathbb{CP}^{N}$ (see \cite{Dias:2010eu,Hoxha:2000jf} for details of geometric features of $\mathbb{CP}^{N}$). Then all deformations on $\mathbb{CP}^{N}$ can be represented as the deformations along the $\theta$ direction. Using this decomposition and the boost property, we take the following metric ansatz  to solve the Einstein-Maxwell equations (\ref{EMeq}) 
%
\begin{eqnarray}
ds^{2} =  -A (e^{(0)})^{2}+2u_{0}e^{(0)}dr -2C_{i}e^{(i)}e^{(0)} 
+G_{ij}e^{(i)}e^{(j)} +r^{2}\sin^{2}{\theta}\,d\Sigma^{2}_{N-1}.
\label{mansatz}
\end{eqnarray}
%
The gauge field ansatz is given by
%
\begin{eqnarray}
A_{\mu}dx^{\mu} = A_{0}e^{(0)} +A_{i}e^{(i)}
\label{gansatz}.
\end{eqnarray}
%
We used the vielbeins, $e^{(0)}$ and $e^{(i)}\,\,(i=1,2,3)$,  defined by
%
\begin{eqnarray}
e^{(0)}=Vd\bar{t},~~e^{(1)}=r\bar{\bPsi},~~e^{(2)}=rd\theta,~~e^{(3)}=r\cot{\theta} \bA.
\end{eqnarray}
%
These ansatz respect the boost property of the Myers-Perry black holes. We set $g_{ir}e^{(i)}dr=0$ as a gauge choice in the metric ansatz. The large $D$ black holes have very large radial gradients with $O(n)$, so we use the radial coordinate $\sR$ defined in eq. (\ref{Rdef}) and set to $r_{0}=1$ as a unit scale. The gradients along other directions are $O(n^{0})$. So we regard metric and gauge fields as functions of $(t,\sR,\psi,\theta)$. The large $D$ scalings of metric and gauge field functions are
%
\begin{eqnarray}
A=O(n^{0}),~~G_{ij}=\delta_{ij}+O(n^{-1}),~~u_{0}=O(n^{0})~~A_{0}=O(n^{0}),
\label{scale1}
\end{eqnarray}
%
and
%
\begin{eqnarray}
C_{i}=O(n^{-1}),~~A_{i}=O(n^{-1}).
\label{scale2}
\end{eqnarray}
%
This scaling assumption comes from the observation of the equally rotating Myers-Perry black holes (\ref{MPsolLD}). The scaling conditions on $G_{ij}$ and $C_{i}$ imply that the $\mathbb{CP}^{N}$ symmetry is preserved at the leading order, and the subleading correction breaks the $\mathbb{CP}^{N}$ symmetry. Hence the leading order solution at large $D$ is regarded as equally rotating solution. We define the large $D$ equally rotating solution by the metric with these scalings. We solve the Einstein-Maxwell equations with a cosmological constant under the ansatz (\ref{mansatz}) and (\ref{gansatz}) for metric and gauge field functions.

\paragraph{Leading order solutions}

At the leading order of the Einstein-Maxwell equations, there are only $\sR$-derivatives in equations. Then we can solve the leading order equations easily and obtain the following simple leading order solutions
%
\begin{eqnarray}
A = 1-\frac{\rho(t,\psi,\theta)}{\sR}+\frac{q(t,\psi,\theta)^{2}}{2\sR^{2}} ,~~u_{0}=V^{-2},~~G_{ij}=\delta_{ij},~~A_{0}=\frac{q(t,\psi,\theta)}{\sR},
\label{LOsol1}
\end{eqnarray}
%
and
%
\begin{eqnarray}
C_{i} = \frac{1}{n}\Biggl[
\frac{p_{i}}{\sR}\left(1-\frac{q^{2}}{2\rho\sR}\right) -2\delta_{i1}\ca\,\sa\,\log{\sR}
\Biggr],~~
A_{i} = -\frac{1}{n}\frac{qp_{i}}{\rho\sR}.
\label{LOsol2}
\end{eqnarray}
%
where $\rho(t,\psi,\theta)$, $q(t,\psi,\theta)$ and $p_{i}(t,\psi,\theta)$ are integration functions in $\sR$-integrations of leading order equations. $V$ is a constant given in eq. (\ref{Vdef}). 
The leading order solutions have horizons at $\sR=\rho_{\pm}$, where $A$ vanishes. $\rho_{\pm}$ is defined by
%
\begin{eqnarray}
\rho_{\pm} = \frac{\rho \pm \sqrt{\rho^{2}-2q^{2}}}{2}.
\end{eqnarray}
%
We can regard $\sR=\rho_{+}$ and $\sR=\rho_{-}$ as an outer and inner horizon of the solution respectively. The leading order solutions have been obtained so that the all metric functions are regular at the outer horizon $\sR=\rho_{+}$. 

In leading order equations we assumed that $u_{0}$ is a constant and obtained eqs. (\ref{LOsol1}) and (\ref{LOsol2}). The redshift factor of the background geometry can be read from the large $\sR$ behavior of $A$, and it is $V$. The fact that the redshift factor is constant is due to the assumption that $u_{0}$ is constant.  If we allow the $(\psi,\theta)$ dependences in $u_{0}$ at the leading order, the leading order solutions are given by eqs. (\ref{LOsol1}) and (\ref{LOsol2}) with $(\psi,\theta)$ dependent redshift factor as studies in \cite{Emparan:2015hwa} for neutral non-rotating black holes. In this paper we consider a constant $u_{0}$ for simplicity. 

\subsection{Effective equations}
At the next to leading order the Einstein-Maxwell equation contains only $\sR$-derivatives for next-to-leading order variables of metric and gauge field functions again, so we can solve them easily. The difference from the leading orders are existence of source terms by leading order solutions. Furthermore there are nontrivial conditions for integration functions in leading order solutions, which come from the constraint equations on $r=\text{constant}$ surfaces
\footnote{
The constraint equation on $r=\text{constant}$ surface are obtained by regarding $r$ direction as a dynamical direction of the system. Hence the constraint equations do not involve the second derivatives with respect to $r$ or equivalently $\sR$. The constraint equations at leading order are satisfied trivially. 
}.  The conditions for integration functions are large $D$ effective equations for charged equally rotating black holes, and they are given by
%
\begin{eqnarray}
\ca\,\pv q -\sa\,\pp q -V^{-1}\cot{\theta} \,\pt q+\frac{p_{2}q}{\rho}\cot{\theta}=0,
\label{Eq1}
\end{eqnarray}
%
%
\begin{eqnarray}
\ca\,\pv \rho -\sa\,\pp \rho -V^{-1}\cot{\theta}\, \pt \rho +p_{2}\cot{\theta}=0,
\label{Eq2}
\end{eqnarray}
%
%
\begin{eqnarray}
&&
\ca\,\pv p_{1} -\sa\,\pp p_{1} -V^{-1}\cot{\theta}\, \pt p_{1} -V^{-2}\sssa\,\pv(\rho_{+}-\rho_{-}) \notag \\
&&~~~~~~
+V^{-2}\ca\ssa\,\pp(\rho_{+}-\rho_{-}) 
+2\cca\,\frac{\rho_{+}p_{3}}{V\rho} 
+\frac{p_{1}p_{2}}{\rho} \cot{\theta} \notag \\
&&~~~~~~
-2\ca\sa\cot{\theta}\,\frac{\rho_{+}p_{2}}{V^{2}\rho} 
+\ca\,\pp\rho -\sa\,\pv\rho \notag \\
&&~~~~~~
+V^{-1}\rho_{-}\cot{\theta}\Biggl[
-\sa\,\pv\left( \frac{p_{2}}{\rho} \right) + \ca\,\pp\left( \frac{p_{2}}{\rho} \right) + \pt\left( \frac{p_{1}}{\rho} \right) 
\Biggr]
=0,
\label{Eq3}
\end{eqnarray}
%
%
\begin{eqnarray}
&&
\ca\,\pv p_{2} -\sa\,\pp p_{2} -V^{-1}\cot{\theta}\, \pt p_{2} +2V^{-1}\cot{\theta}\rho_{-}\pt\left( \frac{p_{2}}{\rho} \right) \notag \\
&&~~~~~~~
+2\ca\sa\tan{\theta}\, p_{3} 
+\left( 1-V^{-2}\ssa \right) \pt\rho+2V^{-1}\sa\tan{\theta}\pp\rho\notag \\
&&~~~~~~~ 
-p_{2} \Biggl[
V^{-1}-\frac{p_{2}}{\rho}\cot{\theta} -\frac{\rho_{+}-\rho_{-}}{V\rho}\cot^{2}{\theta}
\Biggr] =0,
\label{Eq4}
\end{eqnarray}
%
and
%
\begin{eqnarray}
&&
\ca\,\pv p_{3} -\sa\,\pp p_{3} -V^{-1}\cot{\theta}\, \pt p_{3}  
-\text{sech}{\,\alpha}\left( 1-V^{-2}\ssa \right) \pp \rho \notag \\
&&~~~~~
+V^{-1}\cot{\theta}\,\rho_{-} \Biggl[
-\text{sech}{\,\alpha}\pp\left(\frac{p_{2}}{\rho}\right) +\pt \left( \frac{p_{3}}{\rho} \right)
\Biggr]
+p_{2}\cot{\theta}\left( \frac{p_{3}}{\rho} -2\tanh{\alpha} \right) \notag \\
&&~~~~~
+ 2V^{-1}\tanh{\alpha}\cot{\theta}\pt \rho 
-\frac{2p_{3}\rho_{+}}{V\rho}  
=0.
\label{Eq5}
\end{eqnarray}
%
Here we defined
%
\begin{eqnarray}
\pv =\frac{1}{V}\frac{\partial}{\partial t}.
\end{eqnarray}
%
The large $D$ charged equally rotating black holes are solutions of these effective equations. These equations are nonlinear equations, so it is difficult to find analytic solutions in general settings.

\subsection{Stationary solutions}

We study stationary solutions of effective equations (\ref{Eq1}), (\ref{Eq2}), (\ref{Eq3}), (\ref{Eq4}) and (\ref{Eq5}). The stationary solution is defined as solutions of effective equations with Killing vectors $\partial_{t}$ and $\partial_{\psi}$. So, for stationary solutions, we assume the axisymmetry with respect to $\pp$\footnote{This is because  the solution rotates along effectively one direction $\bPsi$, and the rigidity theorem prohibits the inhomogeneous stationary solutions along the rotating direction for such solutions \cite{Hawking:1971vc,Hollands:2006rj}.} as an additional symmetry. The $(dt,\bPsi)$ parts of the leading order metric,  $ds^{2}_{(dt,\bPsi)}$, can be written in a simple form as
%
\begin{eqnarray}
&&
ds^{2}_{(dt,\bPsi)} = -V^{2}\cca\left( 1-\frac{\rho}{\sR} +\frac{q^{2}}{2\sR^{2}} \right)(dt-V^{-1}\tanh{\alpha}\,\bPsi)^{2}  \notag \\
&&~~~~~~~~~~~~~~~~~~~~~~~~~~~~~~~~~~~~~~
+\cca( \bPsi -V\tanh{\alpha}dt)^{2}.
\end{eqnarray}
%
Note that we set $r_{0}=1$ in $\sR$ (see eq. (\ref{Rdef})) as a unit scale. Then we can see that the horizon generating Killing vector $\xi$ is
%
\begin{eqnarray}
\xi = \frac{\partial}{\partial t} +\Omega_{\tH} \frac{\partial}{\partial\psi}.
\label{xidef}
\end{eqnarray}
%
$\Omega_{\tH}$ is the leading order horizon angular velocity at large $D$ given by
%
\begin{eqnarray}
\Omega_{\tH} = V\tanh{\alpha}.
\end{eqnarray}
%
We can calculate the leading order of the surface gravity $\kappa$ at the horizon $\rho=\rho_{+}$ with respect to $\xi$ as
%
\begin{eqnarray}
\kappa &=& -\frac{\partial_{r}(\xi^{\mu}\xi_{\mu})}{2\xi_{r}} \Bigl|_{\rho=\rho_{+}}\notag \\
&=&n\frac{V^{2}}{2\ca}\frac{\rho_{+}-\rho_{-}}{\rho_{+}}.
\label{kappa}
\end{eqnarray}
%
We will see that the surface gravity $\kappa$ is constant for stationary solutions below. The extremal limit of the leading order solution is defined by the limit where the surface gravity vanishes. Then there are two different extremal limits\footnote{For de Sitter case $\hat{\Lambda}>0$, there is an additional extremal limit, which is the Nariai limit $V=0$ as we mentioned. }. One is the extremal limit by the rotation. In this extremal limit the boost parameter $\alpha$ satisfies $\tanh{\alpha}=1$, which means the extremal limit by the rotation is at $a=a_{\text{ext}}$ where
%
\begin{eqnarray}
a^{2}_{\text{ext}}=\frac{1}{1-\hat{\Lambda}}.
\label{aext}
\end{eqnarray}
%
Another extremal limit is by the charge. In this limit the charge $q$ satisfies $\rho_{+}=\rho_{-}$. These extremal limits are not correlated like the extremal limit of the Kerr-Newman solution. 

We can solve the effective equations easily for the stationary solutions. The general stationary solutions of effective equations are obtained by $\rho=e^{P(\theta)}$ and
%
\begin{eqnarray}
q=Qe^{P(\theta)},~~
p_{2}=V^{-1}P'(\theta)e^{P(\theta)},~~
\end{eqnarray}
%
%
\begin{eqnarray}
&&
p_{1}=e^{P(\theta)} \Biggl[ 
\Omega  
+\frac{(a_{+}-a_{-}+\hat{\Lambda}-\ssa)\ca\cot{\theta}}{(1-\hat{\Lambda})\sa}P(\theta) \notag \\
&&~~~~~~~~~~~~~~~~~~~~~~~~~~~~~~~~~~~~~~
+\frac{(a_{+}-a_{-})\ca\cot{\theta}}{2(1-\hat{\Lambda})\sa}P'(\theta)
\Biggr],
\end{eqnarray}
%
%
\begin{eqnarray}
&&
p_{3}=\frac{e^{P(\theta)}\cot{\theta}}{2(1-\hat{\Lambda})\ca\sa}\Biggl[
((a_{-}-a_{+})\cot^{2}{\theta} +\hat{\Lambda} +\ssa)P'(\theta) \notag \\
&&~~~~~~~~~~~~~~~~~~~~~~~~~~~~~~~~~~~~~~~
+(a_{+}-a_{-})\cot{\theta} P''(\theta)
\Biggr].
\end{eqnarray}
%
$Q$ is a constant, and we introduced $a_{\pm}$ defined by
%
\begin{eqnarray}
a_{\pm} = \frac{1\pm\sqrt{1-2Q^{2}}}{2}.
\label{adef}
\end{eqnarray}
%
$\Omega$ in $p_{1}$ is an integration constant. $\Omega$ describes just a trivial $O(1/n)$ perturbation of the boost velocity as $\alpha\rightarrow \alpha +\delta \alpha/n$. So we set to $\Omega=0$. The surface gravity (\ref{kappa}) can be calculated without explicit solution for $P(\theta)$ as
%
\begin{eqnarray}
\kappa = n\frac{V^{2}(a_{+}-a_{-})}{2a_{+}\ca}.
\label{kappadef}
\end{eqnarray}
%
This shows that the surface gravity of stationary solutions is constant. From this expression the extremal solution by the charge is $a_{\pm}=1/2$ or, equivalently, $Q=1/\sqrt{2}$. One of remarkable features of stationary solutions is that the charge distribution $q$ is proportional to the mass distribution $\rho$ with the  constant proportional coefficient $Q$. So the charge is distributed homogeneously, and there is no polarization of the charge on the horizon. This is due to the fact that the equally rotating black holes have the boost symmetry with the constant boost parameter. If the boost symmetry is inhomogeneous on the horizon as singly rotating Myers-Perry black holes \cite{Emparan:2013xia}, the inhomogeneity would lead to inhomogeneous distributions and polarization of the charge on the horizon. Such inhomogeneity might make the horizon dynamics of charged rotating solutions complicate even at the large $D$ limit. This can be seen explicitly for even dimensional equally rotating black holes in Appendix \ref{B}. 

The equation for $P(\theta)$ is obtained from effective equations as
%
\begin{eqnarray}
&&
(a_{+}-a_{-})P'''(\theta)  
-(3(a_{+}-a_{-})\cot^{2}{\theta} +\hat{\Lambda} +\ssa )\tan{\theta}\,P''(\theta) \notag \\
&&~~~~
-((a_{+}-a_{-})(1+2\cos^{2}{\theta})\cot^{2}{\theta}-\cos{2\theta}(\hat{\Lambda}+\ssa))\sec^{2}{\theta}\,P'(\theta)=0.
\label{Peq}
\end{eqnarray}
%
This equation (\ref{Peq}) can be solved by
%
\begin{eqnarray}
P(\theta) = p_{0} + d_{0} \cos^{2}{\theta} + b_{0} (\cos{\theta})^{k(Q,L,\alpha)},
\label{Psol}
\end{eqnarray}
%
where
%
\begin{eqnarray}
k(Q,L,\alpha) = \frac{a_{+}-a_{-}+\hat{\Lambda}+\ssa}{a_{+}-a_{-}}.
\end{eqnarray}
%
$p_{0}$, $d_{0}$ and $b_{0}$ are integration constants. Now we consider stationary deformations of charged equally rotating black holes, and deformations are parametrized by $p_{0}$, $d_{0}$ and $b_{0}$. The solution with $p_{0}=d_{0}=p_{0}=0$ is a non-deformed solution, and it is the charged version of the equally rotating Myers-Perry black hole in the Einstein-Maxwell theory. $p_{0}$ and $d_{0}$ represent trivial deformations, so we set  to $p_{0}=0$ and $d_{0}=0$. In fact $p_{0}$ is a redefinition of $r_{0}$ at $O(1/n)$. $d_{0}$ describes deformations of solution in charged version of Myers-Perry family (see section \ref{3})\footnote{We do not know the exact solutions of charged versions of Myers-Perry black holes with general angular momenta. So the solution with $d_{0}\neq 0$ can be regarded as non-trivial deformed solution. But we do not consider such solutions here.}. $b_{0}$ is an amplitude of nontrivial deformations of the solution out of the Myers-Perry family. $k(Q,L,\alpha)$ is a non-integer in general, so the solution with $b_{0}\neq 0$ is not regular at $\theta=\pi/2$. To have regular solutions with $b_{0}\neq 0$, $k(Q,L,\alpha)$ should be a positive even integer as
%
\begin{eqnarray}
k(Q,L,\alpha) =\ell \equiv 2k_{s},
\label{kscond}
\end{eqnarray}
%
where $k_{s}$ is a positive integer and $\ell$ is a positive even integer. We use $\ell$ instead of $k_{s}$ for convenience in the following section. The condition (\ref{kscond}) can be solved for $a$ by using eq. (\ref{alphadef}) as
%
\begin{eqnarray}
a^{2} &= & a_{\ell}^{2} \notag \\
&\equiv &
\frac{(a_{+}-a_{-})(\ell-1)-\hat{\Lambda}}{(1-\hat{\Lambda})(\ell-2a_{-}(\ell-1 )-\hat{\Lambda})}.
\label{aell}
\end{eqnarray}
%
As we will see in section \ref{3}, charged equally rotating black holes become unstable at $a>a_{\ell}$, and there are static perturbations describing this deformed solution at $a=a_{\ell}$ for each $\ell$.  

Note that deformations with $\ell=2$ is absorbed into one with $d_{0}$ in eq. (\ref{Psol}). So the deformations with $\ell=2$ give trivial deformations. The nontrivial deformation starts at $\ell=4$. 

\paragraph{Thermodynamic quantities}

We can read thermodynamic quantities of stationary solutions with $\rho=1$, $q=Q$ and $p_{i}=0$\footnote{The thermodynamic quantities of stationary solutions with $b_{0}\neq 0$ involve the nontrivial integration on $\mathbb{CP}
^{N}$, so we consider the solution with $b_{0}=0$ here for simplicity. }. The leading order metric is given by
%
\begin{eqnarray}
&&
ds^{2}= -V^{2}\cca\left( 1-\frac{\sR_{0}}{\sR} +\frac{Q^{2}}{2}\frac{\sR^{2}_{0}}{\sR^{2}} \right)(dt-rV^{-1}\tanh{\alpha}\,\bPsi)^{2}  \notag \\
&&~~~~
+\cca( r\bPsi -V\tanh{\alpha}dt)^{2} +2(\ca\,dt-rV^{-1}\sa\,\bPsi)dr +r^{2}d\Sigma^{2}_{N}.
\label{QLOsol}
\end{eqnarray}
%
The leading order gauge field is given by
%
\begin{eqnarray}
A_{\mu}dx^{\mu} = Q\frac{\sR_{0}}{\sR}(V\ca\,dt-\sa\,\bPsi).
\label{QLOsol2}
\end{eqnarray}
%
Here we introduce a parameter $\sR_{0}$. This parameter represents $O(1/n)$ redefinition of $r_{0}$. Actually, when we redefine $r_{0}$ by
%
\begin{eqnarray}
r_{0}\rightarrow r_{0}\left( 1-\frac{\log{b}}{n} \right),
\end{eqnarray}
%
$\sR=(r/r_{0})^{n}$ is changed to 
%
\begin{eqnarray}
\sR\rightarrow b\sR.
\end{eqnarray}
%
This parameter $\sR_{0}$ does not have physical effects, and it is just a scaling relation between horizon radii and mass scale. In the following we explicitly write $r_{0}$, which was set to unity, to show the dimension of thermodynamic quantities. This solution reproduces the large $D$ limit metric of the exact solution (\ref{MPsol}) for neutral equally rotating black hole when $Q=0$ with $\sR_{0}=V^{-2}\cosh^{-2}{\alpha}$. We set $\sR_{0}=V^{-2}\cosh^{-2}{\alpha}$ also for charged solution (\ref{QLOsol}). Then the large $D$ limits of thermodynamic quantities of the charged solution (\ref{QLOsol}) can be read from observations on neutral solution as
%
\begin{eqnarray}
M= \frac{n r_{0}^{n}}{16\pi}\Omega_{n+1},~~
J = \frac{r_{0}^{n+1}\ta \,}{8\pi V}\Omega_{n+1},~~
S=\frac{r_{+}^{n+1}\ca}{4 \pi}\Omega_{n+1},
\label{th1}
\end{eqnarray}
%
for the mass, angular momentum and entropy respectively. $r=r_{+}$ is the horizon position. The temperature $T_{\tH}$ and horizon angular velocity $\Omega_{\tH}$ of eq. (\ref{QLOsol}) are given by
%
\begin{eqnarray}
T_{\tH} = \frac{nV^{2}(a_{+}-a_{-})}{4\pi a_{+} r_{+}\ca},~~\Omega_{\tH}=Vr_{+}^{-1}\ta.
\label{th2Q}
\end{eqnarray}
%
 We set the gravitational constant and Boltzmann constant in $D$ dimensions to be unity. $\Omega_{n+1}$ is a volume of unit $S^{n+1}$. $\Omega_{n+1}$ vanishes at large $D$ exponentially as $O(e^{-n/2})$ \cite{Emparan:2013moa}. To see non-vanishing components of thermodynamic quantities we remain $\Omega_{n+1}$ without taking the large $D$ limit. In similar way we can give the charge $\mathcal{Q}$ of the leading order solution from (\ref{QLOsol2}) as
%
\begin{eqnarray}
\mathcal{Q} &=& \frac{nQr_{0}^{n}}{16\pi V\ca}\Omega_{n+1} \notag \\
&=& \frac{n\sqrt{a_{+}a_{-}}r_{0}^{n}}{8\sqrt{2}\pi V\ca}\Omega_{n+1} .
\end{eqnarray}
%
The horizon radius $r_{+}$ is related with the mass scale $r_{0}$ by
%
\begin{eqnarray}
1- \sR_{0}\left(\frac{r_{0}}{r_{+}}\right)^{n} + \frac{Q^{2}\sR^{2}_{0}}{2}\left(\frac{r_{0}}{r_{+}}\right)^{2n}=0.
\label{rpdef}
\end{eqnarray}
%
This relation can be solved by
%
\begin{eqnarray}
r_{+}^{n} = a_{+}\sR_{0}r_{0}^{n}.
\end{eqnarray}
%
$r_{0}$ and $r_{+}$ becomes same at the large $D$ limit as
%
\begin{eqnarray}
r_{+}=r_{0}+O(1/n).
\end{eqnarray}
%
From these results one can see that thermodynamic quantities of charged rotating black holes satisfy the Smarr formula at large $D$ as
%
\begin{eqnarray}
M=T_{\tH}S +\frac{n}{2}\Omega_{\tH}J +\Phi_{\tH}\mathcal{Q}.
\label{smarr}
\end{eqnarray}
%
The factor $n/2$ comes from the $N=n/2$ equal angular momenta. $\Phi_{\tH}$ is the electric potential on the horizon given by
%
\begin{eqnarray}
\Phi_{\tH}=A_{\mu}\xi^{\mu}\Bigl|_{r=r_{+}} =\frac{VQ}{a_{+}\ca} = \frac{\sqrt{a_{-}}\,V}{\sqrt{2a_{+}}\ca},
\end{eqnarray}
%
$\xi_{\mu}$ is the horizon generating Killing vector defined in eq. (\ref{xidef}). The Smarr formula for black holes with a cosmological constant usually has the volume term of the black hole with the cosmological constant \cite{Kastor:2009wy,Kubiznak:2012wp} in the right hand side of eq. (\ref{smarr}). However, from the fact that the volume of unit $S^{n+1}$ is smaller by $1/n$ factor than its area, the volume term does not contribute to the leading order in the Smarr formula at large $D$. So eq. (\ref{smarr}) does not contain the cosmological constant explicitly. This smallness feature of the volume compared with the area is the universal property of the large $D$ black holes \cite{Emparan:2013moa}.

One may think that we can calculate the thermodynamic Hessian of charged rotating black holes from the thermodynamic quantities defined above, and the ultraspinning surface for charged rotating black holes can be defined. As discussed in \cite{Dias:2010maa,Emparan:2014jca}, although the equally rotating black holes have the ultraspinning instability, the instability exists in perturbations which break all symmetry in $\mathbb{CP}^{N}$. Thus, to define the ultraspinning surface of equally rotating solutions from the thermodynamic Hessian, we should know the general formula of thermodynamic quantities of general charged rotating black hole family. In our analysis we consider only the equally rotating solutions, so, we would not be able to define the ultraspinning surface only from above thermodynamic quantities in principle\footnote{We can give some guess-work on this. We have general deformed stationary solutions with parameters $p_{0}$, $d_{0}$ and $b_{0}$. Especially $d_{0}$ represents the $O(1/n)$ deformations within the charged version of Myers-Perry family. So, roughly, to define the ultraspinning surface, we need $d_{0}$-dependences of thermodynamic quantities. If we obtain them, we can calculate the thermodynamic Hessian for charged unequally rotating black holes with $O(1/n)$ angular momentum differences. Such thermodynamic Hessian is enough to define the ultraspinning surface of the charged equally rotating black holes. However we should be careful to calculate the thermodynamic quantities of solutions with $d_{0}\neq 0$ since they involve estimations of on-trivial integrations on $\mathbb{CP}^{N}$ at the large $D$ limit.}. In section \ref{3}, instead, we can find the ultraspinning instabilities of charged equally rotating black holes directly from the quasinormal modes.    

\paragraph{Kerr-Schild form} Let us rewrite the leading order solutions in the Kerr-Schild form. The leading order solutions are given by
%
\begin{eqnarray}
&&
ds^{2} = -V^{2}d\bar{t}^{2} +2d\bar{t}dr +r^{2}\bar{\bPsi}^{2}  +r^{2}d\Sigma^{2}_{N}
+\left( \frac{\rho}{\sR}-\frac{q^{2}}{2\sR^{2}} \right)d\bar{t}^{2} \notag \\
&&~~~~~~~~~~~~~~~~~~~~~~~~~~~~~~~~~~~~~~~~
- \frac{1}{n}\sum_{i=1,2,3}\frac{p_{i}}{\sR}\left( 1-\frac{q^{2}}{2\rho\sR} \right)d\bar{t}e^{(i)},
\end{eqnarray}
%
and
%
\begin{eqnarray}
A_{\mu}dx^{\mu} = \frac{q}{\sR}d\bar{t} +\frac{1}{n} \sum_{i=1,2,3} \frac{q p_{i}}{\rho\sR}e^{(i)}.
\end{eqnarray}
%
Then the leading order solutions can be written in the Kerr-Schild form as
%
\begin{eqnarray}
ds^{2}= \eta_{\mu\nu}dx^{\mu}dx^{\nu} +\left( \frac{\rho}{\sR}-\frac{q^{2}}{2\sR^{2}} \right)k_{\mu}k_{\nu}dx^{\mu}dx^{\nu},~~
A_{\mu}dx^{\mu} = \frac{q}{\sR}k_{\mu}dx^{\mu},
\end{eqnarray}
%
where $\eta_{\mu\nu}$ is a reference flat background metric given by\footnote{In the usual Kerr-Schild form, the effect of the cosmological constant in the metric can be also written as the deviation from flat metric. Here, instead, we absorbed the effect of the cosmological constant into the flat metric as the reference metric. }
%
\begin{eqnarray}
\eta_{\mu\nu}dx^{\mu}dx^{\nu} =  -V^{2}d\bar{t}^{2} +2Vd\bar{t}dr +r^{2}\bPsi^{2}  +r^{2}d\Sigma^{2}_{N}.
\end{eqnarray}
%
$k_{\mu}$ is a vector field defined by
%
\begin{eqnarray}
k_{\mu}dx^{\mu} = Vd\bar{t} - \frac{1}{n} \sum_{i=1,2,3}\frac{p_{i}}{\rho}e^{(i)}.
\end{eqnarray}
%
The vector $k_{\mu}$ is a null vector in the sense that
%
\begin{eqnarray}
\eta^{\mu\nu}k_{\mu}k_{\nu} =O(1/n^{2}).
\end{eqnarray}
%
Hence the leading order metric at large $D$ can be cast into the Kerr-Schild form. This feature comes directly from our assumption that $O(n^{0})$ parts in the metric have a boost symmetry. To check if the solution with higher order corrections in $1/D$ expansions can be written in the Kerr-Schild form, we should solve more higher order equations of the Einstein-Maxwell equations in $1/D$ expansions. We do not pursue this possibility in this paper, although it would be interesting to check if the charged rotating black hole has the Kerr-Schild form even at higher orders.

\section{Quasinormal modes}
\label{3}

In this section we perform the perturbation analysis of the effective equations to find quasinormal modes of charged equally rotating black holes. We take a solution of effective equations
%
\begin{eqnarray}
\rho=1,~~q=Q,~~ p_{i}=0,
\end{eqnarray}
%
as the background solution of perturbations. This solution corresponds to the stationary solution with $b_{0}=0$ in section \ref{2} and describes (homogeneous) charged equally rotating black holes with a cosmological constant. This background solution has $\mathbb{CP}^{N}$ symmetry, so $\theta$-dependent parts of perturbations can be decomposed by using the charged scalar harmonics $\mathbb{Y}^{\ell m}$ on $\mathbb{CP}^{N}$ \cite{Dias:2010eu}. Then the perturbation of this solution is given by
%
\begin{eqnarray}
\rho = 1+\delta \rho\, e^{-i\omega t+im\psi} \,\mathbb{Y}^{\ell m},~~
q = Q\bigl( 1+\delta q\, e^{-i\omega t+im\psi} \,\mathbb{Y}^{\ell m} \bigr),
\end{eqnarray}
%
and
%
\begin{eqnarray}
p_{1,3} = \delta p_{1,3}\, e^{-i\omega t+im\psi}\, \mathbb{Y}^{\ell m},~~
p_{2} = \delta p_{2} \, e^{-i\omega t+im\psi} \, \pt\mathbb{Y}^{\ell m}.
\end{eqnarray}
%
We already performed $r$-integrations  of the Einstein-Maxwell equations to obtain the effective equations. So the perturbation variables are parameterized by the constant amplitudes $\delta\rho, \delta q$ and $\delta p_{i}$ under the decomposition. The charged scalar harmonics $\mathbb{Y}^{\ell m}$ becomes simple at large $D$ (see Appendix \ref{A}) as
%
\begin{eqnarray}
\mathbb{Y}^{\ell m}(\theta) \propto (\cos{\theta})^{\ell}.
\end{eqnarray}
%
$\ell>0$ is a quantum number on $\mathbb{CP}^{N}$ and satisfies $\ell \geq m$. Using this fact, the perturbation equations become a set of simple linear algebraic equations for $\delta \rho, \delta q, \delta p_{i}$. It would be useful for representation of results to introduce following quantities
%
\begin{eqnarray}
\hat{\omega} =\omega \ca,~~\hat{m}=Vm\sa.
\end{eqnarray}
%
Note that zero rotation limit corresponds to $\sa=0$ and $\hat{m}=0$. The perturbations have two different independent sectors, charge perturbation and gravitational perturbations. The charge perturbation is defined by $\delta q \neq \delta \rho$. The gravitational perturbation is defined by $\delta q = \delta \rho$. The charge perturbation describes the fluctuation with a net charge, and the gravitational perturbation gives density fluctuation. In the following we give results for two sectors separately. 

\paragraph{Charge perturbation}

For the charge perturbation the perturbation equations can be solved if the frequency satisfies $\hat{\omega}=\hat{\omega}_{c}$ where
%
\begin{eqnarray}
\hat{\omega}_{c} =  \hat{m} -i\ell.
\end{eqnarray}
%
Hence the charge perturbation is stable. This quasinormal mode for charge perturbations reproduces one of Reissner-Nordstrom black holes in (Anti) de Sitter \cite{Bhattacharyya:2015fdk,Tanabe:2015isb}.

\paragraph{Gravitational perturbation}

The gravitational perturbation has two subsectors: One is the vector type gravitational perturbation defined by $\delta\rho=0$, and another is the scalar type gravitational perturbation defined by $\delta \rho\neq 0$. 

\subparagraph{Vector type gravitational perturbation}

The vector type gravitational perturbation has the decoupled quasinormal mode frequencies $\hat{\omega}=\hat{\omega}_{v}$ given by
%
\begin{eqnarray}
\hat{\omega}_{v} = \hat{m} - ia_{+}\ell.
\end{eqnarray}
%
$a_{\pm}$ was introduced in eq. (\ref{adef}).  So the vector type gravitational perturbation is stable. At the zero rotation limit the frequency $\hat{\omega}_{v}$ reproduces the decoupled quasinormal mode of the vector type gravitational perturbations of the Reissner-Nordstrom black hole in (Anti) de Sitter \cite{Bhattacharyya:2015fdk,Tanabe:2015isb}\footnote{In the notation of \cite{Bhattacharyya:2015fdk,Tanabe:2015isb}, $a_{+}=1/(1+Q^{2})$ and $a_{-}=Q^{2}/(1+Q^{2})$. }. So we call this mode the vector type gravitational perturbation. As we will see below, the scalar type gravitational perturbation contains rotating version of the scalar type gravitational perturbation of the Reissner-Nordstrom black holes in (Anti) de Sitter. 

\subparagraph{Scalar type gravitational perturbation}

The quasinormal mode condition to solve the perturbation equations for scalar type gravitational perturbations is obtained as
%
\begin{eqnarray}
&&
2\hat{\omega}^{3} +2i\hat{\omega}^{2}\Bigl[
3i\hat{m} +a_{+}(3\ell-4)
\Bigr]
+2\hat{\omega}\Bigl[
4-8a_{-}+4a_{-}^{2}-7\ell +14\ell a_{-} -6\ell a_{-}^{2} +3\ell^{2}-6\ell^{2}a_{-} \notag \\
&&~~~~
+2a_{-}^{2}\ell^{2} +6ia_{+}\hat{m}(\ell-1) -3\hat{m}^{2} +(4-\ell)\ssa -\hL(\ell+4\ssa)
\Bigr]
+2ia_{+}\hat{m}^{2}(3\ell-2) \notag \\
&&~~~~
-2\hat{m}^{3} +2ia_{+}\ell(\ell-2)(2a_{-}(\ell-1)-\ell+\Lambda) 
+2\hat{m}(\ell(3\ell-4) +2a_{-}^{2}(\ell-1)(\ell-2) \notag \\
&&~~~~
-2a_{-}(\ell-1)(3\ell-2)-(\ell-2)\hat{\Lambda})
+\cca\,(2ia_{+}\ell(\ell-2) +\hat{m}(4-2\ell-8\hat{\Lambda}))=0.
\label{QNMcondS}
\end{eqnarray}
%
Note that the neutral version ($a_{+}=1$ and $a_{-}=0$) of eq. (\ref{QNMcondS}) is reduced to the quasinormal mode condition of the equally rotating (Anti) de Sitter Myers-Perry black hole \cite{Emparan:2014jca}\footnote{To compare quasinormal mode condition in \cite{Emparan:2014jca} we use the relation $\hat{\Lambda}=-1/L^{2}$. $L$ is a curvature scale of a cosmological constant used in \cite{Emparan:2014jca}}.  At zero rotation limit ($\sa=0$ and $\hat{m}=0$) this condition can be solved by
%
\begin{eqnarray}
\hat{\omega}_{0}= -ia_{+}(\ell-2),
\label{vec}
\end{eqnarray}
%
and
%
\begin{eqnarray}
\hat{\omega}_{\pm} = -ia_{+}(\ell-1)  \pm \sqrt{(a_{+}^{2} - a_{-}^{2}\ell)(\ell-1)+\ell\hat{\Lambda}}.
\label{sca}
\end{eqnarray}
%
These are decoupled quasinormal modes of scalar type gravitational perturbations $(\hat{\omega}_{\pm})$ and vector type gravitational perturbations $(\hat{\omega}_{0})$ of (Anti) de Sitter Reissner-Nordstrom black holes \cite{Bhattacharyya:2015fdk,Tanabe:2015isb}. As discussed in \cite{Emparan:2014jca}, the vector harmonics on $S^{D-2}$ with the quantum number $\ell_{v}$ can be decomposed by two charged scalar harmonics on $\mathbb{CP}^{N}$. The quantum numbers of the corresponding harmonics on $\mathbb{CP}^{N}$ are $\ell=\ell_{v}\pm 1$. So there are two vector type gravitational perturbations on $S^{D-2}$ as $\hat{\omega}_{v}$ and $\hat{\omega}_{0}$ at zero rotation limit. 

\begin{figure}[t]
 \begin{center}
  \includegraphics[width=70mm,angle=0]{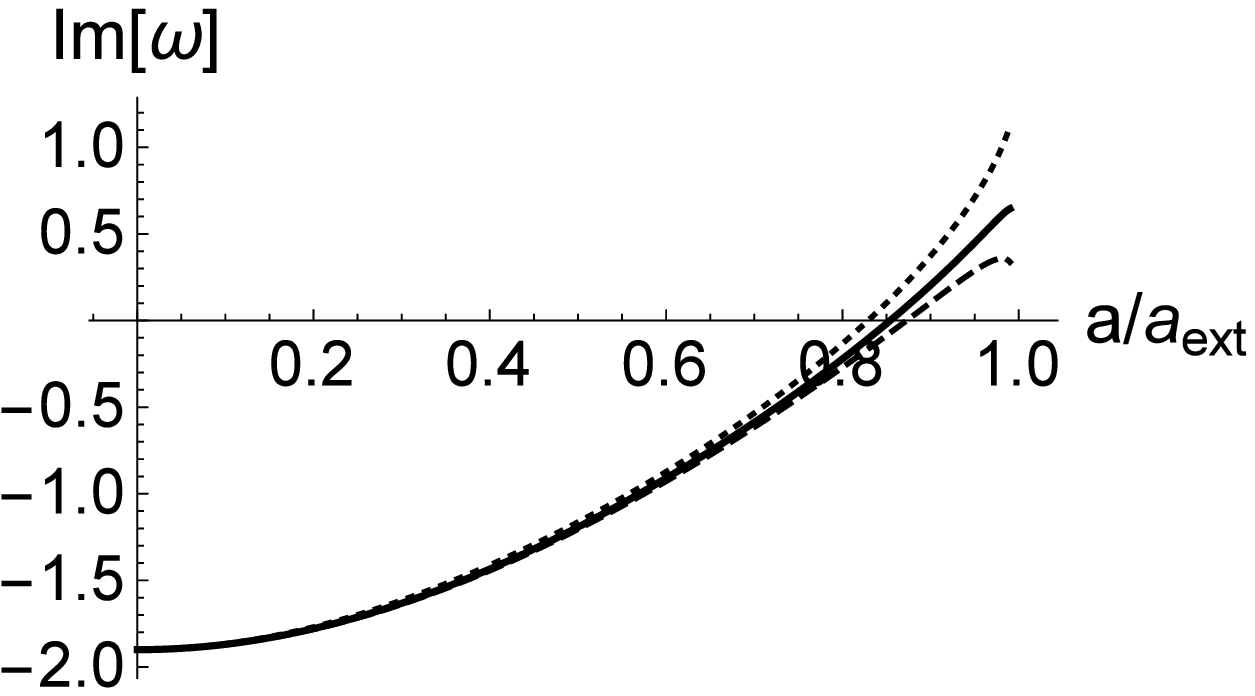}
  \hspace{5mm}
  \includegraphics[width=70mm,angle=0]{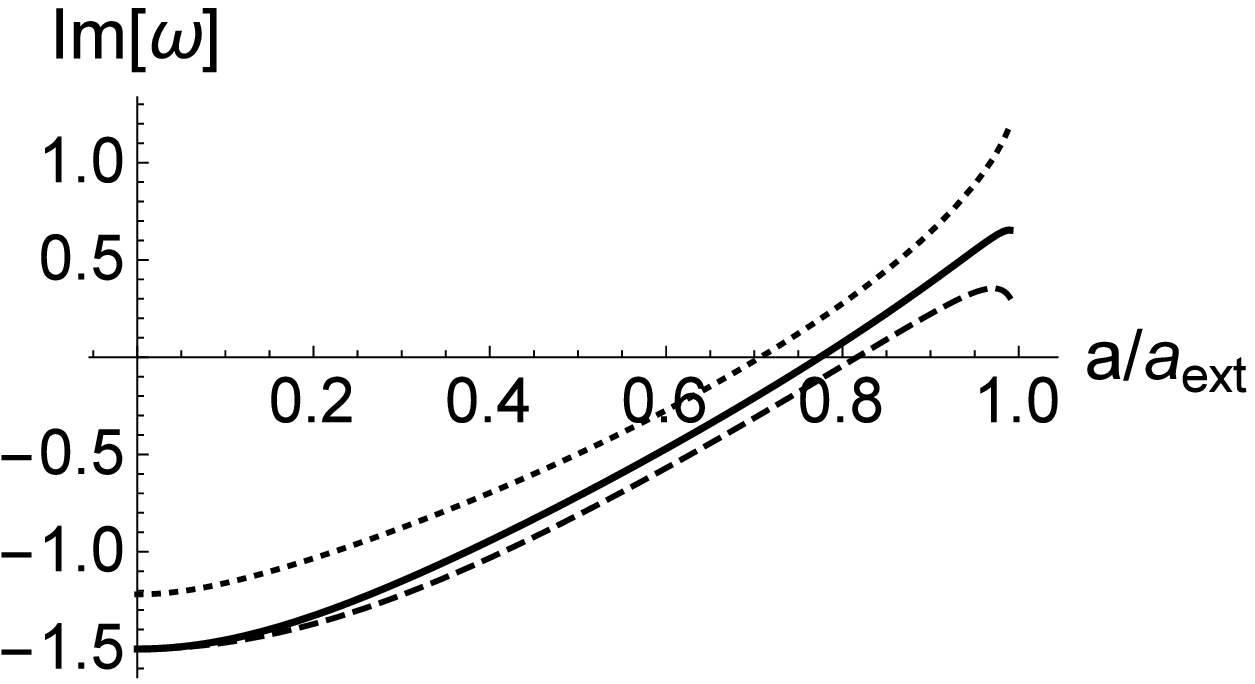}
 \end{center}
 \vspace{-5mm}
 \caption{  
The plots of the quasinormal modes of the unstable mode in the scalar type gravitational perturbations with $\ell=4$ and $m=0$ are shown for $\hat{\Lambda}=0$ (thick line), $\hat{\Lambda}=0.5$ (dotted line) and $\hat{\Lambda}=-0.5$ (dashed line). The left panel is at $a_{-}=0.05$, and the right panel is at $a_{-}=0.25$.  }
 \label{fig:ell4m0}
\end{figure}

One of interesting features of scalar type gravitational perturbations is the existence of unstable modes. We can see that the frequency satisfies the superradiance condition
%
\begin{eqnarray}
\omega \bigl|_{a=a_{\ell}}= m\Omega_{\tH} = m V\tanh{\alpha},
\label{super}
\end{eqnarray}
%
or equivalently $\hat{\omega}=\hat{m}$, at the critical rotation $\alpha=\alpha_{\ell}$ where $\alpha_{\ell}$  is given by 
%
\begin{eqnarray}
\sinh^{2}{\alpha_{\ell}} =(a_{+}-a_{-})(\ell-1)-\hat{\Lambda}.
\label{alphaell}
\end{eqnarray}
%
Thus the frequency becomes purely real for non-axisymmetric perturbation and zero for axisymmetric perturbation at the critical rotation. Using eq. (\ref{alphadef}), this condition for $\alpha$ can be rewritten to the condition for the rotation parameter $a$ as $a=a_{\ell}$. $a_{\ell}$ was given in eq. (\ref{aell}) and satisfies
%
\begin{eqnarray}
a^{2}_{\ell} = \frac{(a_{+}-a_{-})(\ell-1)-\hat{\Lambda}}{(1+\hat{\Lambda})(\ell -2a_{-}(\ell-1) +\hat{\Lambda})}<a^{2}_{\text{ext}}.
\end{eqnarray}
%
$a_{\text{ext}}$ is the rotation parameter at the extremal limit (see eq. (\ref{aext})). At larger rotation than the critical rotation $a>a_{\ell}$ one of gravitational perturbations becomes unstable. So gravitational perturbations become unstable in subextremal region. For axisymmetric perturbations $(m=0)$ there are stationary zero mode perturbations at $a=a_{\ell}$ (see eq. (\ref{super})). The existence of stationary zero mode perturbations implies that there is a new solution branch at $a=a_{\ell}$. This new solution branch corresponds to the deformed stationary solution constructed in section \ref{2}. $b_{0}$ in eq. (\ref{Psol}) is a parameter of the new solution branches. In Figure \ref{fig:ell4m0} we show the plots of unstable mode in scalar type gravitational perturbations with $\ell=4$ and $m=0$ for $\hat{\Lambda}=0$ (thick line), $\hat{\Lambda}=0.5$ (dotted line) and $\hat{\Lambda}=-0.5$ (dashed line). The left panel is at $a_{-}=0.05$, and the right panel is at $a_{-}=0.25$.  
%
%
We can see that the quasinormal mode shows the instability at the subextremal region. The unstable mode for $a_{-}=0.05$ is originated from $\hat{\omega}_{0}$ in eq. (\ref{vec}), while the unstable mode for $a_{-}=0.25$ has its origin in $\hat{\omega}_{0}$ for $\hat{\Lambda}=0$ and $\hat{\Lambda}=-0.5$ and in $\hat{\omega}_{+}$ (see eq. (\ref{sca})) for $\hat{\Lambda}=0.5$. Actually, as observed in \cite{Tanabe:2015isb}, the quasinormal mode of the scalar type gravitational perturbations of Reissner-Nordstrom black holes in de Sitter becomes purely imaginary at larger charge than the certain charge. Our result implies that such mode becomes unstable when we add the rotation.    

The most unstable modes are $\ell=2$ modes. The quantum number $m$ should satisfy $\ell \geq m$, so the most unstable modes are $(\ell,m)=(2,2)$ and $(2,1)$ at the leading order in $1/D$ expansions\footnote{Including $1/n$ correction we can see the difference between $(\ell,m)=(2,2)$ and $(2,1)$ modes for neutral solutions \cite{Emparan:2014jca}. Then we observed that the most unstable mode is  $(\ell,m)=(2,2)$. We expect that the same thing would happen, and the most unstable mode is  $(\ell,m)=(2,2)$ also for charged rotating solutions when we include $O(1/n)$  corrections.}. As for $(\ell,m)=(2,0)$ mode, the quasinormal mode condition (\ref{QNMcondS}) can be solved explicitly by
%
\begin{eqnarray}
\hat{\omega}_{0}^{(2,0)}=0,~~ \hat{\omega}^{(2,0)}_{\pm}= \pm \sqrt{1-2a_{-}-a_{-}^{2}+2\ssa +2(1+2\ssa)\hat{\Lambda}}-ia_{+}.
\end{eqnarray}
%
$\hat{\omega}^{(2,0)}_{0}$ corresponds to the mode with $(\ell,m)=(2,0)$ which shows the behavior in eq. (\ref{super}). But this mode is a marginally unstable mode, and it describes just the deformations in Myers-Perry family \cite{Dias:2010eu,Emparan:2014jca}. Hence $\hat{\omega}_{0}^{(2,0)}$ is not a physical perturbation. The stationary mode $\hat{\omega}_{0}^{(2,0)}$ describes the linearized deformation with respect to $d_{0}$ in the stationary solution (\ref{Psol}). This can be understood from $\mathbb{Y}^{\ell=2,m=0}\propto \cos^{2}{\theta}$. Thus $d_{0}$ in eq. (\ref{Psol}) is not a physical deformation parameter.

The instabilities in gravitational perturbations can be seen clearly for asymptotically flat case $(\hat{\Lambda}=0)$. The quasinormal mode condition (\ref{QNMcondS}) can be solved explicitly for the modes with $\ell=m$ by
%
\begin{figure}[t]
 \begin{center}
  \includegraphics[width=70mm,angle=0]{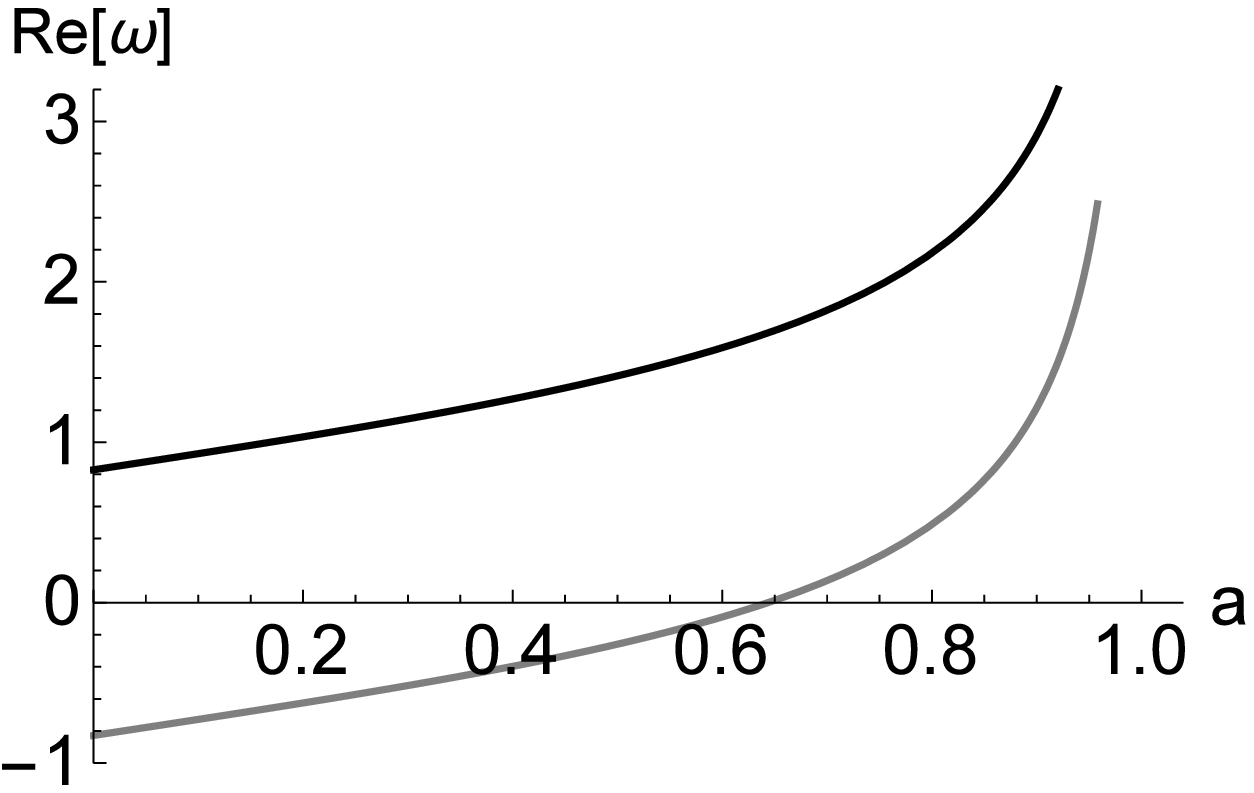}
  \hspace{5mm}
  \includegraphics[width=70mm,angle=0]{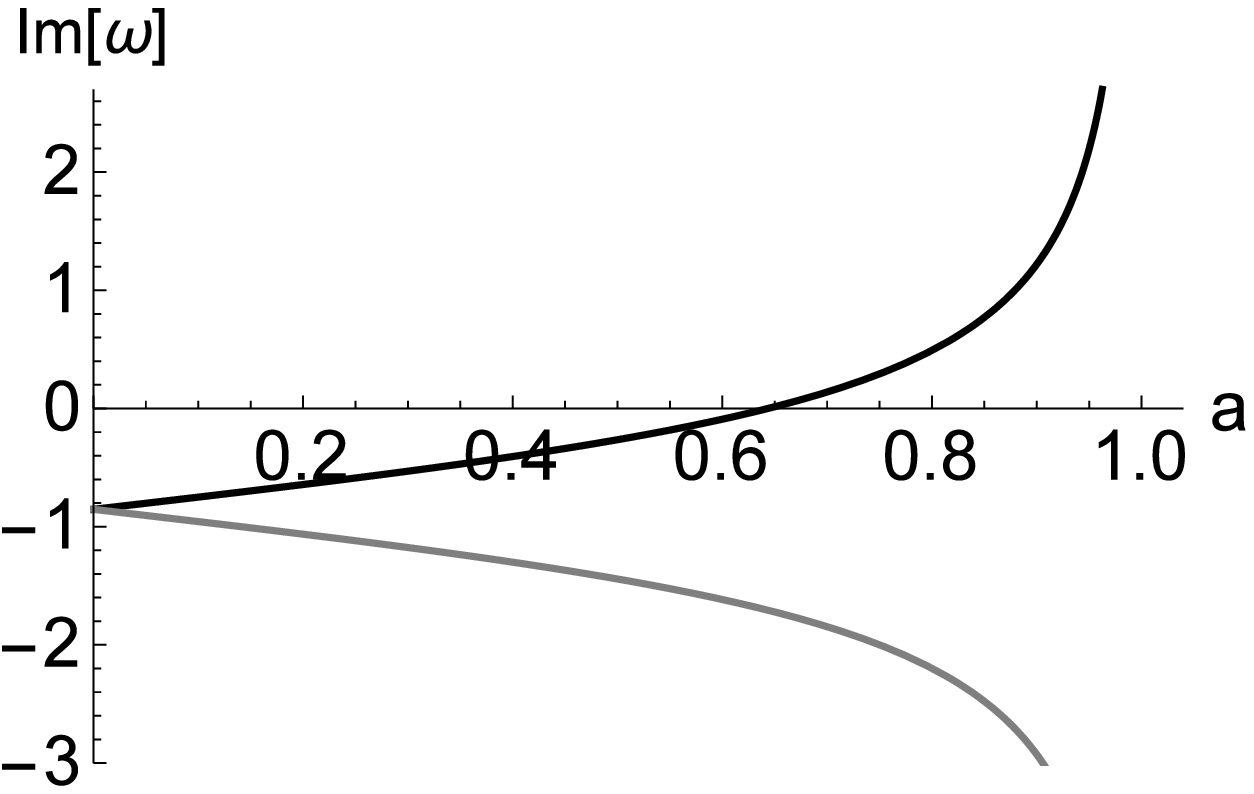}
 \end{center}
 \vspace{-5mm}
 \caption{  
The plots of $\omega^{\ell=m}_{\pm}$ with $\ell=m=2$ at $a_{-}=0.25$ are shown. The left (right) panel shows the real (imaginary) part. The black (gray) line is for $\omega^{\ell=m}_{+}$ ($\omega^{\ell=m}_{-}$) .}
 \label{fig:ellm2}
\end{figure}
%
%
\begin{eqnarray}
&&
\omega^{\ell=m}_{\pm}=(m-1)\ta -i\frac{a_{+}(m-1)}{\ca} \notag \\
&&~~~~~~~
\pm \frac{\sqrt{(m-1)(a^{2}_{-}m+a_{+}^{2}+2ia_{+}\sa -\ssa)}}{\ca}.
\end{eqnarray}
%
These quasinormal modes are shown in Figure \ref{fig:ellm2}. The black line (gray line) corresponds to $\omega^{\ell=m}_{+}$ ($\omega^{\ell=m}_{-}$) with $\ell=m=2$ mode at $a_{-}=0.25$ . We can see that $\omega_{+}$ shows the instability at $\alpha>\alpha_{\ell}$. Note that the quasinormal mode condition (\ref{QNMcondS}) is a cubic algebraic equation for $\omega$. So there are three roots. But one of them would not be a physical mode for $\ell=m$ mode perturbations. This was shown for neutral solutions \cite{Emparan:2014jca}, and we expect that the same situation holds for the charged rotating case. So we show only two modes $\omega_{\pm}$ for $\ell=m$ mode perturbations.

Let us observe the charge effects on the condition for the critical rotation (\ref{alphaell}). The critical rotation for charged solution becomes smaller than for neutral solution as
%
\begin{eqnarray}
\alpha_{\ell}(Q=0)>\alpha_{\ell}(Q\neq 0).
\label{QAcond}
\end{eqnarray}
%
This means that the charge of black holes can help the ultraspinning instability of equally rotating Myers-Perry black holes. This is consistent with the results in \cite{Caldarelli:2010xz}, which observed that nearly extremal solution by the charge can have the blackfold description without large angular momenta. The enough large charge can cancel the tension of the brane with a small rotation, and, then, such nearly extremal solutions can be analyzed by the blackfold approach. So their results indicate that the nearly extremal solution by the charge would be unstable by the Gregory-Laflamme instability of the black brane even at a small rotation, and eq. (\ref{QAcond}) confirms their results\footnote{Our solution is an equally rotating one in odd dimensions. This solution was not included in their analysis \cite{Caldarelli:2010xz} since the horizon cannot be extended infinitely like singly rotating black holes. However, for equally rotating solutions, we can define the ultraspinning solutions by using the Hessian of thermodynamic quantities \cite{Dias:2010eu}. So we expect that the statement in \cite{Caldarelli:2010xz} would hold for equally rotating solutions.}.  

%
\begin{figure}[t]
 \begin{center}
  \includegraphics[width=80mm,angle=0]{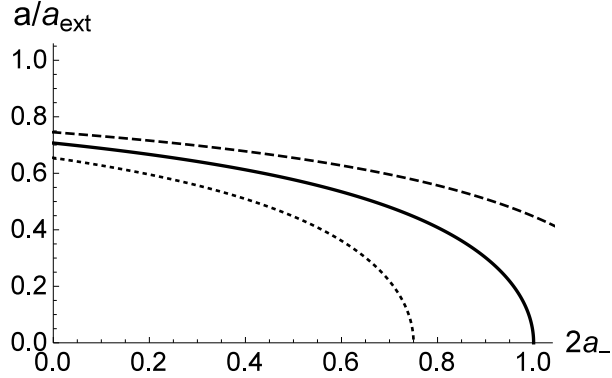}
 \end{center}
 \vspace{-5mm}
 \caption{  The plots of the critical rotation of $\ell=2$ mode perturbations for de Sitter with $\Lambda=0.25$ (dotted line), Anti de Sitter with $\hat{\Lambda}=-0.25$ (dashed line) and asymptotically flat case (thick line) are shown. The region above each lines is unstable region.}
 \label{fig:aQ}
\end{figure}
%
For asymptotically flat case ($\hat{\Lambda}=0$) the critical rotation for extremal solution by charge $(a_{+}=a_{-}=1/2)$ becomes zero. This may imply that extremal Reissner-Nordstrom black hole is marginally unstable against gravitational perturbations. However it is not clear that our large $D$ analysis can be applied to extremal solutions since the extremal solution has a vanishing surface gravity while the large $D$ method for black holes assumes that the surface gravity is $O(D)$. To treat extremal solutions by large $D$ expansion method, we need more careful analysis. 

For asymptotically de Sitter case $(\hat{\Lambda}>0)$, we find that the critical rotation becomes zero for subextremal solution which satisfies
%
\begin{eqnarray}
(a_{+}-a_{-})(\ell-1)=\hat{\Lambda}.
\label{Qcritical}
\end{eqnarray}
%
This means that the non-rotating solution, that is, Reissner-Nordstrom black hole in de Sitter, is unstable. Actually the condition (\ref{Qcritical}) is equivalent to the condition for the threshold charge of the instability of de Sitter Reissner-Nordstrom black holes at large $D$ \cite{Tanabe:2015isb}. Thus the ultraspinning instability of equally rotating Myers-Perry black hole is connected with the instability of de Sitter Reissner-Nordstrom black holes in rotation-charge plane of the solution parameter space. This implies that the origins of instabilities of Myers-Perry black hole and Reissner-Nordstrom black hole in de Sitter are same. 

For asymptotically Anti de Sitter case $(\hat{\Lambda}<0)$, the critical rotation is non-zero even at the extremal limit by charge. This means the stability of Anti de Sitter Reissner-Nordstrom black holes, and this is consistent with numerical results \cite{Konoplya:2008rq}. 

These situations can be seen in Figure \ref{fig:aQ}. In Figure \ref{fig:aQ} we plot the $Q$ dependence of the critical rotation  (\ref{aell}) for the most unstable $\ell=2$ mode of de Sitter with $\hat{\Lambda}=0.25$ (dotted line), Anti de Sitter with $\hat{\Lambda}=-0.25$ (dashed line) and  asymptotically flat case (thick line). The region above each lines in $(a_{-},a)$ plane is the unstable region of each solutions. As we can see the asymptotically de Sitter solution (dotted line) has an unstable region even at zero rotation limit $(a=0)$. 

\section{Summary}
\label{4}

We have considered large $D$ effective theory for charged equally rotating black holes in odd dimensions. The important point is the fact that charged equally rotating black holes in odd dimensions have the Kerr-Schild form, at least, at the large $D$ limit. This Kerr-Schild form of the metric makes the analysis much simpler. For example, the presence of the Kerr-Schild form implies that the metric at the large $D$ limit should have the boost symmetry. Then, the rotation can be introduced simply by the boost transformation of non-rotating solutions. As a result we can solve the Einstein-Maxwell equations for charged equally rotating black holes and obtain the effective equations. This simplification of the analysis at large $D$ might not happen for charged singly rotating black holes or charged equally rotating black holes in even dimensions. The effective equations describe the dynamical nonlinear evolution of black holes. The perturbative analysis of effective equations give quasinormal modes of charged rotating black holes with/without a cosmological constant. The gravitational perturbations are unstable at larger rotation or larger charge. Especially the charged equally rotating black hole in de Sitter can be unstable even at zero rotation limit. This instability has been observed as one of de Sitter-Reissner-Nordstrom black holes \cite{Konoplya:2008au,Cardoso:2010rz,Konoplya:2013sba,Tanabe:2015isb}. So this result suggests that  the ultraspinning instability and instability of de Sitter Reissner-Nordstrom black holes have same origin as dynamical phenomena. This was already pointed out in the blackfold analysis \cite{Caldarelli:2010xz}, and it was mentioned that the electric charge can play a similar role with the rotation. Then the small rotation with enough charge can be treated by the blackfold analysis, and it implies that such black hole would have Gregory-Laflamme like instability.  Our results in this paper confirmed this possibility by observing quasinormal modes directly. 

There are some interesting extensions of our current work. One is to extend our analysis to other theories such as Einstein-Maxwell-dilaton theory and Einstein-Maxwell theory with Chern-Simons term. In some of these theories, we have exact solutions of charged rotating black holes \cite{Breckenridge:1996is,Breckenridge:1996sn,Llatas:1996gh,Kunz:2006jd}. So it is possible to perform the stability analysis of exact solutions by effective equations. Another is to consider charged singly rotating black holes in the Einstein-Maxwell theory. In our analysis the Kerr-Schild form of the metric at the large $D$ limit is a crucial property to obtain the effective equations. It is not clear, however, that the all charged rotating black holes can have the Kerr-Schild form at the large $D$ limit. In some analysis negative results for this possibility was concluded \cite{Myers:1986un,Aliev:2006yk}. If the Kerr-Schild form is a special feature of equally rotating solutions in the Einstein-Maxwell theory, we may think that physical properties of charged equally rotating solution in odd dimensions are not same with other charged rotating black holes. To check this it would be interesting to investigate charged singly rotating black holes.   

\section*{Acknowledgments}

This work was supported by JSPS Grant-in-Aid for Scientific Research No.26-3387.

\appendix

\section{Harmonics on $\mathbb{CP}^{N}$ at large $D$}
\label{A}

In this appendix we give large $D$ behaviors of charged scalar harmonics $\mathbb{Y}^{\ell m}$ on $\mathbb{CP}^{N}$. The charged scalar harmonics $\mathbb{Y}^{\ell m}$ on $\mathbb{CP}^{N}$ is defined by the following eigenvalue equation on $\mathbb{CP}^{N}$ \cite{Dias:2010eu,Hoxha:2000jf}
%
\begin{eqnarray}
(\hat{\nabla}^{2}+\lambda)\mathbb{Y}^{\ell m}=0,
\label{Ydef}
\end{eqnarray}
%
where $\hat{\nabla}$ is the gauge-covariant derivative given by
%
\begin{eqnarray}
\hat{\nabla}=\nabla -im\bA.
\end{eqnarray}
%
$\nabla$ is the covariant derivative with respect to the Fubini-Study metric on $\mathbb{CP}^{N}$. The eigenvalue $\lambda$ is parametrized by integers $\ell$ and $m$ as
%
\begin{eqnarray}
\lambda = \ell(\ell+2N)-m^{2}.
\end{eqnarray}
%
The Fubini-Study metric $d\Sigma^{2}_{N}$ on $\mathbb{CP}^{N}$ can be decomposed by using $d\Sigma^{2}_{N-1}$ on $\mathbb{CP}^{N-1}$ as
%
\begin{eqnarray}
d\Sigma^{2}_{N} = d\theta^{2} +\cot^{2}{\theta}\bA^{2} +d\Sigma^{2}_{N-1}.
\end{eqnarray}
%
$\theta$ is one of non-Killing coordinates on $\mathbb{CP}^{N}$. The charged scalar harmonics $\mathbb{Y}^{\ell m}$ on $\mathbb{CP}^{N}$ can be also decomposed 
%
\begin{eqnarray}
\mathbb{Y}^{\ell m} = f(\theta)\hat{\mathbb{Y}},
\end{eqnarray}
%
where $\hat{\mathbb{Y}}$ is the charged scalar harmonics on $\mathbb{CP}^{N-1}$. Let us consider the charged scalar harmonics $\mathbb{Y}^{\ell m}$ without excitations on $\mathbb{CP}^{N-1}$, which means $\hat{\mathbb{Y}}$ is constant. Then the eigenvalue equation (\ref{Ydef}) becomes
%
\begin{eqnarray}
f''(\theta) +(2N+1)f'(\theta) +(\ell(\ell+2N)-m^{2}-m^{2}\cos^{2}{\theta}) f(\theta)=0.
\label{feq}
\end{eqnarray}
%
Note that $D=n+3=2N+3$ dimensional equally rotating Myers-Perry black holes have $\mathbb{CP}^{N}$ in the geometry. So the large $D$ limit corresponds to the large $N$ limit. Then, taking the large $N$ limit of eq. (\ref{feq}), we obtain the explicit solution for $f(\theta)$ at large $D$ as
%
\begin{eqnarray}
\mathbb{Y}^{\ell m} = f_{0} (\cos{\theta})^{\ell} +O(1/n).
\end{eqnarray}
%
$f_{0}$ is an integration constant, which gives a normalization of $\mathbb{Y}^{\ell m}$. The charged scalar harmonics has a simple form $\mathbb{Y}^{\ell m} \propto (\cos{\theta})^{\ell}$ at large $D$, and there is no $m$-dependence.

\section{Equally rotating black holes with a small charge in even dimensions}
\label{B}

In this appendix we give large $D$ analysis on equally rotating black holes with a small charge in even dimensions. We consider the Einstein-Maxwell equation with a cosmological constant
%
\begin{eqnarray}
R_{\mu\nu}-\frac{1}{2}Rg_{\mu\nu}+\Lambda g_{\mu\nu} = \frac{1}{2}\left( 
F_{\mu\rho}F^{\rho}{}_{\nu}-\frac{1}{4}F_{\rho\sigma}F^{\rho\sigma}g_{\mu\nu}
\right),~~\nabla^{\mu}F_{\mu\nu}=0,
\label{EMeqB}
\end{eqnarray}
%
in $D=2N+4$ dimensional spacetime. As done for odd dimensional case in section \ref{2}, let us observe the exact solution of the neutral equally rotating solution \cite{Myers:1986un,Gibbons:2004js,Gibbons:2004uw} in even $(D=2N+4)$ dimensions. The metric in ingoing Eddington-Finkelstein coordinates is\footnote{Our radial coordinate, $r$, is related with the radial coordinate $r_{\text{BL}}$ in the Boyer-Lindquist coordinate \cite{Myers:1986un,Gibbons:2004js,Gibbons:2004uw} by $r^{2}=(r^{2}_{\text{BL}}+a^{2})$. Furthermore we choose $\psi$ coordinate so that the metric component $g_{t\Psi}\bPsi dt$ vanishes at large $r$. }
%
\begin{eqnarray}
&&
ds^{2}= \frac{2}{(1+a^{2}\hat{\Lambda})\sqrt{r^{2}-a^{2}}}\left( (1+a^{2}\hat{\Lambda}\cos^{2}{z})dt -ar\sin^{2}{z}\bPsi \right) dr \notag \\
&&
-\frac{1+a^{2}\hat{\Lambda}\cos^{2}{z}}{1+a^{2}\hat{\Lambda}}\left( 1-(r^{2}-a^{2})\hat{\Lambda}
-\frac{\sqrt{r^{2}-a^{2}}}{r^{2}-a^{2}\sin^{2}{z}}\left(\frac{r_{0}}{r}\right)^{n} \right) dt^{2} \notag \\
&&
+\frac{r^{2}\sin^{2}{z}}{1+a^{2}\hat{\Lambda}}\left( 1+\frac{a^{2}\sin^{2}{z}\sqrt{r^{2}-a^{2}}}{r^{2}(1+a^{2}\hat{\Lambda})(r^{2}-a^{2}\sin^{2}{z})}\left(\frac{r_{0}}{r}\right)^{n}  \right) \bPsi^{2} \notag \\
&&
-\frac{2a\sin^{2}{z}\sqrt{r^{2}-a^{2}}(1+a^{2}\hat{\Lambda}\cos^{2}{z})}{(1+a^{2}\hat{\Lambda})^{2}(r^{2}-a^{2}\sin^{2}{z})}\left(\frac{r_{0}}{r}\right)^{n}  \bPsi dt \notag \\
&&
+\frac{r^{2}-a^{2}\sin^{2}{z}}{1+a^{2}\hat{\Lambda}\cos^{2}{z}}dz^{2} 
+\frac{r^{2}\sin^{2}{z}}{1+a^{2}\hat{\Lambda}}d\Sigma^{2}_{N},
\label{soleven}
\end{eqnarray}
%
where $n=2N$\footnote{In this appendix we use $1/n$ as an expansion parameter to sole the Einstein-Maxwell equations instead of $1/D$. } . $\hat{\Lambda}$ is defined by
%
\begin{eqnarray}
\Lambda =\frac{(n+2)(n+3)}{2}\hat{\Lambda}.
\end{eqnarray}
%
$\bPsi$ is the $1$-form on $\mathbb{CP}^{N}$ given by
%
\begin{eqnarray}
\bPsi= d\psi +\bA.
\end{eqnarray}
%
$d\Sigma^{2}_{N}$ is the Fubini-Study metric on $\mathbb{CP}^{N}$. $a$ is a rotation parameter, and $r_{0}$ is a parameter for the horizon radius. $\bA$ is the $1$-form on $\mathbb{CP}^{N}$, and the definition is same with one used for odd dimensions in section \ref{2}. 

The $(dt,dr,\bPsi)$ part of the solution has the boost symmetry at the large $D$ limit. The $(dt,dr,\bPsi)$ part of the metric (\ref{soleven}) becomes
%
\begin{eqnarray}
ds^{2}_{(dt,dr,\bPsi)} = \eta_{AB}dx^{A}dx^{B} +\frac{\sR_{0}}{\sR}u_{A}u_{B}dx^{A}dx^{B} +\frac{2V(z)}{\hat{\kappa}\ca}u_{A}dx^{A}dr 
\end{eqnarray}
%
at large $D$. $\sR$ is defined by
%
\begin{eqnarray}
\sR=\left( \frac{r}{r_{0}} \right)^{n},
\end{eqnarray}
%
and we set to $r_{0}=1+O(1/n)$. $\sR_{0}$ and $\hat{\kappa}$ are given by
%
\begin{eqnarray}
\sR_{0} = \frac{\sqrt{1-a^{2}}}{1-(1-a^{2})\hat{\Lambda}},~~
\hat{\kappa}=\sqrt{1-a^{2}}(1-(1-a^{2})\hat{\Lambda}).
\end{eqnarray}
%
$\eta_{AB}$ is a two dimensional flat metric with Lorentzian signature. $dx^{A}=(V(z)dt, G(z)\bPsi)$, and functions $V(z)$ and $G(z)$ are given by
%
\begin{eqnarray}
V(z)^{2} = \frac{(1-(1-a^{2})\hat{\Lambda})(1+a^{2}\hat{\Lambda}\cos^{2}(z))}{1+a^{2}\hat{\Lambda}},~~
G(z)^{2} = \frac{\sin^{2}{z}}{1+a^{2}\hat{\Lambda}}.
\end{eqnarray}
%
$u_{A}dx^{A}=V(z)d\bar{t}$ is the velocity field of the solution at large $D$. The rest frame $d\bar{x}^{A}=(V(z)d\bar{t}, G(z)\bar{\bPsi})$ is given by
%
\begin{eqnarray}
&&
d\bar{t}= \ca\, dt -V(z)^{-1}G(z)\sa\,\bPsi\\
&&
\bar{\bPsi}= \ca\,\bPsi -V(z)G(z)^{-1}\sa\,dt.
\end{eqnarray}
%
The crucial difference from the odd dimensional solution is that the boost parameter $\alpha$ is not constant. The boost parameter $\alpha$ is given by
%
\begin{eqnarray}
\ta = a\sin{z}\sqrt{\frac{1-(1-a^{2})\hat{\Lambda}}{1+a^{2}\hat{\Lambda}\cos^{2}{z}}}.
\end{eqnarray}
%
The even dimensional equally rotating Myers-Perry black hole in (Anti) de Sitter can be obtained by the boost transformation from the (Anti) de Sitter Schwarzschild black hole at the large $D$ limit, but its boost transformation is not homogeneous like the odd dimensional solutions\footnote{To reproduce precisely the large $D$ limit of the (Anti) de Sitter Myers-Perry black hole from the boost transformation of the (Anti) de Sitter Schwarzschild metric, we need rescaling of $z$ coordinate and the Fubini-Study metric $d\Sigma^{2}_{N}$. But such rescaling is not crucial in the following arguments and results. }. This is similar with singly rotating Myers-Perry black holes \cite{Emparan:2013xia}. This inhomogeneous boost property of the large $D$ metric, however, is still useful. For example, we can obtain quasinormal mode frequencies of the singly rotating Myers-Perry black holes analytically by using the (inhomogeneous) boost symmetry \cite{Suzuki:2015iha}. 

We solve the Einstein-Maxwell equation (\ref{EMeqB}) for equally rotating black holes with a small charge by using the boost symmetry at the large $D$ limit. The metric ansatz is
%
\begin{eqnarray}
ds^{2}= -A(e^{(0)})^{2} + 2u_{0}e^{(0)}dr -C_{i}e^{(0)}e^{(i)}+G_{ij}e^{(i)}e^{(j)}+\frac{r^{2}\sin^{2}{z}\sin^{2}{\theta}}{1+a^{2}\hat{\Lambda}}d\Sigma^{2}_{N-1},
\end{eqnarray}
%
and the gauge field ansatz is
%
\begin{eqnarray}
A_{\mu}dx^{\mu}= A_{0}e^{(0)} +A_{i}e^{(i)}.
\end{eqnarray}
%
We used the decomposition (\ref{CPN}) of the metric on $\mathbb{CP}^{N}$ and introduced the coordinate $\theta$. The vielbeins, $e^{(0)}$ and $e^{(i)} \,(i=1,2,3,4)$, are defined by
%
\begin{eqnarray}
e^{(0)}=V(z)d\bar{t},~~
e^{(1)}=rG(z)\bar{\bPsi},~~
e^{(2)}=\sqrt{\frac{r^{2}-a^{2}\sin^{2}{z}}{1+a^{2}\hat{\Lambda}\cos^{2}{\theta}}}dz,~~
\end{eqnarray}
%
and
%
\begin{eqnarray}
e^{(3)}=\frac{r\sin{z}\cot{\theta}}{\sqrt{1+a^{2}\hat{\Lambda}}} \bA,~~
e^{(4)}=\frac{r\sin{z}}{\sqrt{1+a^{2}\hat{\Lambda}}}d\theta.
\end{eqnarray}
%
These ansatz respect the boost symmetry of exact solutions (\ref{soleven}) at the large $D$ limit. 
The metric and gauge field functions are functions of $(t,\sR,z,\psi,\theta)$. 

The large $D$ behavior of metric and gauge field functions are
%
\begin{eqnarray}
A=O(n^{0}),~~u_{0}=O(n^{0}),~~C_{i}=O(n^{-1}),~~G_{ij}=\delta_{ij}+O(n^{-1}),
\end{eqnarray}
%
and
%
\begin{eqnarray}
A_{0}=O(n^{-1/2}),~~A_{i} =O(n^{-3/2}). 
\end{eqnarray}
%
The factor $O(n^{-1/2})$ in the large $n$ behaviors of gauge field functions implies that we consider solutions with a small charge, and its charge is $O(n^{-1/2})$.  

\paragraph{Leading order solutions}
The leading orders of the Einstein-Maxwell equations at large $D$ contain only $\sR$-derivatives. Then we can easily obtain the leading order solutions by integrating them with respect to $\sR$. The results are
%
\begin{eqnarray}
A=1-\frac{\rho}{\sR},~~
u_{0}= \frac{V(z)}{\hat{\kappa}\ca},~~
C_{i}=\frac{1}{n}\frac{p_{i}}{\sR},~~
G_{ij}=\delta_{ij}+O(1/n^{2}),
\end{eqnarray}
%
and
%
\begin{eqnarray}
A_{0}=\frac{1}{\sqrt{n}}\frac{q}{\sR},~~A_{i}=\frac{1}{n\sqrt{n}}\frac{q p_{i}}{\rho \sR}. 
\end{eqnarray}
%
$\rho(u,z,\theta,\psi)$, $q(u,z,\theta,\psi)$ and $p_{i}(u,z,\theta,\psi)$ are integration functions in leading order solutions. They describe the mass, charge and momentum density respectively.

\paragraph{Effective equations}
At the next to leading orders of the Einstein-maxwell equations, we obtain the constraint equations for integration functions in leading order equations. The constraint equations compose effective equations for even dimensional equally rotating black holes with a small charge at the large $D$ limit. The effective equations are
%
\begin{eqnarray}
&&
\mathcal{L}q+ q\Biggl[
\frac{a^{2}(1+a^{2}\hL\cos^{2}{z})\ca\,\cos^{2}{z}}{\sqrt{1-a^{2}}(1-a^{2}\sin^{2}{z})} \notag \\
&&~~~~~~~~~~
+\frac{(1+a^{2}\hL)V(z)\cot{\theta}}{\sin^{2}{z}}\frac{p_{4}}{\rho}
+\frac{V(z)(1+a^{2}\hL\cos^{2}{z})\cot{z}}{1-a^{2}\sin^{2}{z}}\frac{p_{2}}{\rho}
\Biggr]=0,
\label{eq1B}
\end{eqnarray}
%
%
\begin{eqnarray}
\mathcal{L}{\rho} +\frac{V(z)(1+a^{2}\hL)\cot{\theta}}{\sin^{2}{z}}p_{4}+\frac{V(z)(1+a^{2}\hL\cos^{2}{z})\cot{z}}{1-a^{2}\sin^{2}{z}}p_{2}=0,
\label{eq2B}
\end{eqnarray}
%
%
\begin{eqnarray}
&&
\mathcal{L}p_{1} -\frac{1-a^{2}(1+\cos^{2}{z})}{1-a^{2}\sin^{2}{z}}\sa\,\partial_{t} \rho +\frac{2(1+a^{2}\hL \cos^{2}{z})\cot{z}\sa}{\sqrt{1-a^{2}}(1-a^{2}\sin^{2}{z})}\pz \rho \notag \\
&&~~~~
+\frac{V(z)(1+a^{2}\hL)(1-a^{2}(1+\cos^{2}{z}))\ca\,G(z)}{(1-a^{2})\sin^{2}{z}(1-a^{2}\sin^{2}{z})}\pp \rho\notag \\
&&~~~~
+p_{1}\Biggl[
\frac{(1+a^{2}(1+\hL)+a^{4}\hL -a^{2}(1-(2-a^{2})\hL)\cos^{2}{z})\ca\,\cot^{2}{z}}{\sqrt{1-a^{2}}(1-a^{2}\sin^{2}{z})} \notag \\
&&~~~~~~~~
+\frac{V(z)(1+a^{2}\hL)\cot{\theta}}{\sin^{2}{z}}\frac{p_{4}}{\rho} 
+\frac{V(z)(1+a^{2}\hL\cos^{2}{z})\cot{z}}{1-a^{2}\sin^{2}{z}}\frac{p_{2}}{\rho}
\Biggr] \notag \\
&&~~~~
-\frac{2V(z)(1+a^{2}\hL)(2-(1-a^{2})\hL +a^{2}\hL\cos^{2}{z})\ca\sa\cot{z}}{(1-(1-a^{2})\hL)(1-a^{2}\sin^{2}{z})}p_{2} \notag \\
&&~~~~
+\frac{2(1+a^{2}\hL\cos^{2}{z})\ca}{\sqrt{1-a^{2}}\sin^{2}{z}}p_{3}
-\frac{2V(z)(1+a^{2}\hL)^{2}\ca\sa\cot{\theta}}{(1-(1-a^{2})\hL)\sin^{2}{z}}p_{4}=0,
\label{eq3B}
\end{eqnarray}
%
%
\begin{eqnarray}
&&
\mathcal{L}p_{2} +\frac{V(z)a^{2}\sin{2z}}{\sqrt{1-a^{2}}(1-a^{2}\sin^{2}{z})(1+a^{2}\hL\cos^{2}{z})}\partial_{t}\rho
+\frac{V(z)(1-a^{2}(1+\cos^{2}{z}))}{1-a^{2}\sin^{2}{z}}\pz\rho \notag \\
&&~~~~
-\frac{2G(z)(1+a^{2\hL})\ca\sa\cot{z}}{\sqrt{1-a^{2}}\sin^{2}{z}}\pp\rho
+\frac{2V(z)(1+a^{2}\hL)\ca\sa\cot{z}}{1+a^{2}\hL\cos^{2}{z}}p_{1} \notag \\
&&~~~~
-p_{2}\Biggl[
\frac{(1+a^{2}\hL\cos^{2}{z})(1-a^{2}-2\cos^{2}{z}+a^{2}\cos^{4}{z})\ca}{\sqrt{1-a^{2}}(1-a^{2}\sin^{2}{z})\sin^{2}{z}}\notag \\
&&~~~~~~~~~~~~~
-\frac{V(z)(1+a^{2}\hL\cos^{2}{z})\cot{z}}{1-a^{2}\sin^{2}{z}}\frac{p_{2}}{\rho} 
-\frac{V(z)(1+a^{2}\hL)\cot{\theta}}{\sin^{2}{z}}\frac{p_{4}}{\rho}
\Biggr] \notag \\
&&~~~~
-\frac{2(1+a^{2}\hL)\ca\cot{z}\cot{\theta}}{\sqrt{1-a^{2}}\sin^{2}{z}}p_{4}
-\frac{V(z)\sin{2z}}{1-a^{2}\sin^{2}{z}}\frac{q^{2}}{\rho}=0,
\label{eq4B}
\end{eqnarray}
%
%
\begin{eqnarray}
&&
\mathcal{L}p_{3} -\frac{V(z)G(z)(1-2a^{2})(1+a^{2}\hL)^{2}(1-a^{2}\sin^{2}{z})\ca}{(1-a^{2})\sin^{2}{z}(1+a^{2}\hL\cos^{2}{z})}\pp\rho \notag \\
&&~~~~
-p_{3}\Biggl[
\frac{(1+a^{2}\hL)(2-2a^{2}-(1-2a^{2})\cos^{2}{z})\ca}{\sqrt{1-a^{2}}\sin^{2}{z}} \notag \\
&&~~~~~~~~~~~
-\frac{V(z)(1+a^{2}\hL\cos^{2}{z})\cot{z}}{1-a^{2}\sin^{2}{z}}\frac{p_{2}}{\rho} 
-\frac{V(z)(1+a^{2}\hL)\cot{\theta}}{\sin^{2}{z}}\frac{p_{4}}{\rho}
\Biggr] \notag \\
&&~~~~
-\frac{2(1+a^{2}\hL)(1-a^{2}\sin^{2}{z})\cot{\theta}\sa}{\sqrt{1-a^{2}}\sin^{2}{z}}\pt\rho \notag \\
&&~~~~
-\frac{2V(z)(1+a^{2}\hL)^{2}(1-a^{2}\sin^{2}{z})\ca\sa\cot{\theta}}{\sin^{2}{z}(1+a^{2}\hL\cos^{2}{z})}p_{4}=0,
\label{eq5B}
\end{eqnarray}
%
and
%
\begin{eqnarray}
&&
\mathcal{L}p_{4} +\frac{2G(z)(1+a^{2}\hL)(1-a^{2}\sin^{2}{z})\ca\sa\tan{\theta}}{\sqrt{1-a^{2}}\sin^{2}{z}}\pp\rho
 \notag \\
&&~~~~
+\frac{V(z)(1-2a^{2})}{1-a^{2}}\pt\rho
+2V(z)\ca\sa\tan{\theta}p_{3} \notag \\
&&~~~~
+p_{4}\Biggl[
\frac{V(z)(1+a^{2}\hL\cos^{2}{z})\cot{z}}{1-a^{2}\sin^{2}{z}}\frac{p_{2}}{\rho}
+\frac{V(z)(1+a^{2}\hL)\cot{\theta}}{\sin^{2}{z}}\frac{p_{4}}{\rho}
\Biggr] \notag \\
&&~~~~
+\frac{\ca p_{4}}{2\sqrt{1-a^{2}}\sin^{2}{z}\sin^{2}{\theta}(1-a^{2}\sin^{2}{z})}\Bigl[
2(1-a^{2})^{2}(1+a^{2}\hL)\cos{2z} \notag \\
&&~~~~~~~~
+(2-a^{2}-(2-5a^{2}+4a^{4}(1-\hL)+4a^{6}\hL)\cos{2z})\cos^{2}{z} \notag \\
&&~~~~~~~~
+a^{2}(1+(2-a^{2})\hL -(1+2\hL -2a^{4}\hL -a^{2}(2+\hL))\cos{2z})\cos^{4}{z} \notag \\
&&~~~~~~~~
+2a^{4}\hL\sin^{2}{z}\cos^{6}{z}
\Bigr]=0.
\label{eq6B}
\end{eqnarray}
%
Here we introduced the differential operator $\mathcal{L}$ defined by
%
\begin{eqnarray}
&&
\mathcal{L} = \ca\frac{\partial}{\partial t} -\frac{(1+a^{2}\hat{\Lambda}\cos^{2}{z})\ca\,\cot{z}}{\sqrt{1-a^{2}}}\frac{\partial}{\partial z} \notag \\
&&~~
-\frac{(1+a^{2}\hat{\Lambda})V(z)G(z)\sa}{\sin^{2}{z}}\frac{\partial}{\partial \psi}
-\frac{(1+a^{2}\hat{\Lambda})(1-a^{2}\sin^{2}{z})\ca\,\cot{\theta}}{\sqrt{1-a^{2}}}\frac{\partial}{\partial \theta}.  
\label{Ldef}
\end{eqnarray}
%
The solutions of effective equations describe the equally rotating black holes with a small charge in even dimensions.  

\paragraph{Stationary solutions}
The effective equations have a solution
%
\begin{eqnarray}
\rho=1,~~q=0,~~p_{i}=0.
\end{eqnarray}
%
This neutral solution corresponds to the equally rotating Myers-Perry black hole in even dimensions. To see the charge effects on this solution, we try to find the stationary solution of effective equations by the following ansatz
%
\begin{eqnarray}
\rho =1,~~q=q(z),~~p_{i}=p_{i}(z). 
\label{stationary}
\end{eqnarray}
%
In this solution we do not include $\theta$ dependences in the solution. So the solution satisfying this ansatz has the $\mathbb{CP}^{N}$ symmetry, and it can be regarded as the equally rotating black holes with a small charge. One important observation is that the effective equation (\ref{eq4B}) contain the term $\propto q^{2}$. So the stationary solution can describe the backreaction of the charge fields. Note that we do not intend to find general stationary solutions of effective equations as done in odd dimension case in section \ref{2}. The general stationary solution would depend on $z$ and $\theta$. So the effective equations for general stationary solutions are still partial differential equation, and it would be difficult to solve them.

The ansatz (\ref{stationary}) gives a solution as
%
\begin{eqnarray}
q(z) = \frac{Q\ca}{\sqrt{1+a^{2}\hL\cos^{2}{z}}},~~
p_{2}=-\frac{Q^{2}\ca\sa}{1-(1-a^{2})\hL},
\label{chargesol}
\end{eqnarray}
%
and $p_{1,3,4}=0$. The leading order solution has the horizon at $\sR=\rho$. The stationary solution we are considering has a constant density distribution because the mass density is  constant as $\rho=1$. However the charge distribution becomes inhomogeneous on this solution  as seen in eq. (\ref{chargesol}). The odd dimensional general stationary solution has always homogeneous charge distribution (see section \ref{2}). The inhomogeneity of the charge distribution is characterized by the boost velocity $\propto \ca$\footnote{The factor $1+a^{2}\hL\cos^{2}{z}$ is an effect of the background geometry. The inhomogeneity by this factor arises also by the rotation effects.}. The boost velocity is not homogeneous on the horizon, and such inhomogeneity leads the inhomogeneous charge distribution. Furthermore this charge distribution makes the polarization of the charge on the horizon, and, as a result, the momentum $p_{2}$ is excited as eq. (\ref{chargesol}). For the odd dimensional solution, if we consider non-deformed solution (stationary solution with $b_{0}=0$ in section \ref{2}), there is no excited momentum. From this fact, we expect that the charged rotating solution with inhomogeneous boost velocity become more complicate than the solution with homogeneous boost velocity such as odd dimensional equally rotating solution. This is because, if we consider $O(1)$ charge which means $Q=O(\sqrt{n})$ in eq. (\ref{chargesol}), $p_{2}$ becomes large as $O(n)$. We are assuming the momentum along the inhomogeneous $z$-direction is $O(1/n)$ for solutions with a small charge, and its coefficient is given by $p_{2}$. Now, so, we should consider $O(1)$ momentum along the inhomogeneous $z$-direction for solutions with $O(1)$ charge. In order to study such solution we need more involved analysis since we have considered only the solution which has $O(1)$ momenta along the rotating direction. The charged solutions with inhomogeneous boost velocity would have $O(1)$ momenta not only along the rotating direction, but also the inhomogeneous direction of solutions. Such analysis is not easy even at the large $D$ limit, and it would be one of reasons why we do not have exact charged rotating solutions of the Einstein-Maxwell theory in a simple expression.

\paragraph{Quasinormal modes}

Finally we perform the stability analysis of the neutral stationary solution given by
%
\begin{eqnarray}
\rho=1,~~q=0,~~p_{i}=0.
\end{eqnarray}
%
Note that this stationary solution has the horizon generating Killing vector $\xi$
%
\begin{eqnarray}
\xi =\frac{\partial}{\partial t} +\Omega_{\tH}\frac{\partial}{\partial \psi},
\end{eqnarray}
%
where $\Omega_{\tH}$ is the leading order horizon angular velocity given by
%
\begin{eqnarray}
\Omega_{\tH} =a(1-(1-a^{2})\hL).
\label{omegaB}
\end{eqnarray}
%
The surface gravity is calculated as
%
\begin{eqnarray}
\kappa &=& \frac{n\hat{\kappa}}{2} \notag \\
&=& \frac{n\sqrt{1-a^{2}}(1-(1-a^{2})\hL)}{2}.
\end{eqnarray}
%
So the extremal limit by the rotation is $a=1$. For asymptotically de Sitter case ($\hL>0$),  $\hL=1/(1-a^{2})$ is the Narial limit. For Anti de Sitter case $\hL<0$, there is a rigid upper bound on the rotation for regular solutions \cite{Chrusciel:2006zs} as
%
\begin{eqnarray}
1+a^{2}\hL > 0.
\label{aLB}
\end{eqnarray}
%

The perturbation of the stationary solution is given by
%
\begin{eqnarray}
\begin{array}{c}
\rho =1+\epsilon e^{-i\omega t+im\psi}F_{\rho}(z)\mathbb{Y}^{\ell m},~~
q =\epsilon e^{-i\omega t+im\psi}F_{q}(z)\mathbb{Y}^{\ell m},~\\
p_{1,2,4} =\epsilon e^{-i\omega t+im\psi}F_{1,2,4}(z)\mathbb{Y}^{\ell m},~~
p_{3} =\epsilon e^{-i\omega t+im\psi}F_{3}(z)\pt\mathbb{Y}^{\ell m}.
\end{array}
\end{eqnarray}
%
$\epsilon$ is the perturbation parameter. $\mathbb{Y}^{\ell m}$ is the charged scalar harmonics, and we can set to  $\mathbb{Y}^{\ell m}=(\cos{\theta})^{\ell}$. The differential operator $\mathcal{L}$ in eq (\ref{Ldef}) has a pole at $\cos{z}=0$\footnote{From the fact $G(z)\sa\propto \sin^{2}{z}$, there is no pole at $\sin{z}=0$.}. So we should impose the regularity condition for perturbation variables $F_{\rho}, F_{q}$ and $F_{i}$ at $z=\pi/2$. As the regularity condition to solve the perturbation equations, we impose the following regularity condition
%
\begin{eqnarray}
F_{\rho}(z)\Bigl|_{z=\pi/2}\propto (\cos{z})^{j},~~
F_{q}(z)\Bigl|_{z=\pi/2}\propto (\cos{z})^{j},~~
F_{1}(z)\Bigl|_{z=\pi/2}\propto (\cos{z})^{j-1},
\end{eqnarray}
%
with a positive integer $j$. The behaviors of $F_{i}$ can be obtained from effective equations. 
By imposing this regularity condition, we can obtain the quasinormal modes of the equally rotating Myers-Pery black hole in even dimensions at large $D$. To show results, it is useful to introduce the following quantities
%
\begin{eqnarray}
V_{\Lambda}=1-(1-a^{2})\hL,~~
\hat{\omega}=\frac{V_{\Lambda}}{\sqrt{1-a^{2}}}\omega,~~
\hat{m}=\frac{a}{\sqrt{1-a^{2}}}m,
\end{eqnarray}
%
and
%
\begin{eqnarray}
\hat{\ell}=(1-a^{2}V_{\Lambda})\ell,~~
J=j+\hat{\ell}.
\end{eqnarray}
%
The perturbation is classified into two sectors: charge perturbation $(F_{q}(z)\neq 0)$ and gravitational perturbation $(F_{q}(z)=0)$. 

\subparagraph{Charge perturbation}
The charge perturbation has the quasinormal mode frequency $\omega=\omega_{c}$ where
%
\begin{eqnarray}
\omega_{c}= \frac{\hat{m}}{\sqrt{1-a^{2}}}-i\frac{J}{\sqrt{1-a^{2}}}.
\end{eqnarray}
%
So the charge perturbation is stable.

\subparagraph{Gravitational perturbation}

The gravitational perturbation has two subsectors: One is the vector type gravitational perturbation $(F_{\rho}(z)=0)$, and another is the scalar type gravitational perturbation $(F_{\rho}(z)\neq 0)$. 

The vector type gravitational perturbation is given by $F_{\rho}(z)=0$ and $F_{q}(z)=0$. Then, from  effective equations, we can see $F_{1,3,4}(z)=0$. Thus there is only one perturbation variables, $F_{2}(z)$. We impose the regularity condition on $F_{2}(z)$ by
%
\begin{eqnarray}
F_{2}(z)\Bigl|_{z=\pi/2}\propto (\cos{z})^{j},
\end{eqnarray}
%
with a positive integer $j$. Then we find that the vector type gravitational perturbation has the quasinormal mode frequency $\omega=\omega_{v}$ given by
%
\begin{eqnarray}
\omega_{v}= \frac{\hat{m}}{\sqrt{1-a^{2}}}-i\frac{J}{\sqrt{1-a^{2}}}.
\label{vQNMB}
\end{eqnarray}
%
This is the same frequency with charge perturbations, and the vector type gravitational perturbation is stable. 

The scalar type gravitational perturbation should have the frequency which satisfies
%
\begin{eqnarray}
&&
\hat{\omega}^{4} 
+2i\hat{\omega}^{3}\Bigl[
2J-3+2i\hat{m} +2a^{2}V_{\Lambda}
\Bigr] 
+\hat{\omega}^{2}\Bigl[
-6J^{2} +J(18-12i\hat{m} -(1+12a^{2})V_{\Lambda}) \notag \\
&&~~~~
+2(3\hat{m}^{2}-6 +a^{2}(8+\hat{\ell} -2V_{\Lambda} )V_{\Lambda} -i\hat{m}(5a^{2}V_{\Lambda}-8 ))
\Bigr] 
+2\hat{\omega}\Bigl[
-2\hat{m}^{3}+\hat{m}(8+6J^{2}  \notag \\
&&~~~~
+(1-12a^{2}-2a^{2}\hat{\ell} +3a^{2}V_{\Lambda})V_{\Lambda} +i\hat{m}(4a^{2}V_{\Lambda}-7)+J(6i\hat{m}-16+V_{\Lambda}+10a^{2}V_{\Lambda})) \notag \\
&&~~~~
-i(2J^{3}-4+J^{2}(V_{\Lambda}-9+6a^{2}V_{\Lambda}) +a^{2}V_{\Lambda}(8+4\hat{\ell} -4V_{\Lambda} -3\hat{\ell}V_{\Lambda}) \notag \\
&&~~~~
 +2JV_{\Lambda}(6- 
(1+a^{2}(8+\hat{\ell})V_{\Lambda} +3a^{2}V_{\Lambda})))
\Bigr]
+J^{4}+\hat{m}^{4}-2i\hat{m}^{3}(a^{2}V_{\Lambda}-2)  \notag \\
&&~~~~
+J^{2}(4i\hat{m}-6+V_{\Lambda}+4a^{2}V_{\Lambda}) 
 -4a^{2}\hat{\ell}V_{\Lambda}(2-V_{\Lambda})(2-V{\Lambda})
-2\hat{m}^{2}(2 \notag \\
&&~~~~
+(1-a^{2}(4+\hat{\ell}))V_{\Lambda} +a^{2}V_{\Lambda}^{2}) -2J^{2}(3\hat{m}^{2}-6 
+(2+a^{2}(8+\hat{\ell}))V_{\Lambda} -4a^{2}V_{\Lambda}^{2} \notag \\
&&~~~~
 +i\hat{m}(8-V_{\Lambda}-5a^{2}V_{\Lambda}) )
-2i\hat{m}V_{\Lambda}(2 
-a^{2}(4+4\hat{\ell} -2V_{\Lambda} -3\hat{\ell}V_{\Lambda}))
+J(-4i\hat{m}^{3} \notag \\
&&~~~~
+\hat{m}^{2}(14-V_{\Lambda} -8a^{2}V_{\Lambda})  
+2i\hat{m}(8^(1+2a^{2}(6+\hat{\ell}))V_{\Lambda} +5a^{2}V_{\Lambda}^{2})
+2(-4 \notag \\
&&~~~~
+(2+4a^{2}(2+\hat{\ell}))V_{\Lambda} 
-a^{2}(8+3\hat{\ell})V_{\Lambda}^{2}+2a^{2}V_{\Lambda}^{3}))=0.
\label{sQNMB}
\end{eqnarray}
%
This quasinormal mode condition is complicate but shows simple and interesting feature of scalar type gravitational perturbations. In fact eq. (\ref{sQNMB}) can be solved for $j=0$ modes by
%
\begin{eqnarray}
\omega = m\Omega_{\tH}
\end{eqnarray}
%
or equivalently $\hat{\omega}=\hat{m}$, at the critical rotation $a=a_{c}$ where 
%
\begin{eqnarray}
a_{c}^{2}=\frac{1}{2}\left( \frac{\ell+1}{\ell}-\hL^{-1} +\frac{\sqrt{(\ell-\hL-\ell\hL)^{2}-4\ell\hL(1-\ell-\hL)}}{\ell\hL} \right).
\label{acB}
\end{eqnarray}
%
$\Omega_{\tH}$ was given in eq. (\ref{omegaB}). Note that $a_{c}$ in eq. (\ref{acB}) is finite at $\hL=0$. Actually, we can see
%
\begin{eqnarray}
a_{c}\Bigl|_{\Lambda=0}=\frac{\ell-1}{\ell}.
\end{eqnarray}
%
At larger rotation than the critical rotation $a>a_{c}$ one of scalar type gravitational perturbations becomes unstable. So, for axisymmetric perturbation $(m=0)$, there is a stationary perturbation at $a=a_{c}$. This instability is the ultraspinning instabilities of the equally rotating Myers-Perry black holes. The mode with $j>0$ does not show any instabilities\footnote{Note that eq. (\ref{sQNMB}) is a quartic equation for $\omega$, so there are four modes. But one of them should be a gauge mode. This can be understood by counting the number of decoupled quasinormal modes. The Schwarzschild black hole has three decoupled quasinormal mode, one vector mode and two scalar modes on $S^{D-2}$ \cite{Emparan:2014aba}. Furthermore the vector mode on $S^{D-2}$ splits into two scalar modes on $\mathbb{CP}^{N}$ \cite{Emparan:2014jca}. So there should be four decoupled quasinormal modes in equally rotating Myers-Perry black holes in even dimensions. One of them has been found already as the vector type gravitational perturbation in eq. (\ref{vQNMB}). Thus the scalar type gravitational perturbation should have only three modes. Hence one of solutions for eq. (\ref{sQNMB}) should not be a physical mode.   }. This is consistent with the analysis for the violation of the Breitenlohner-Freedman bound of perturbations at the near horizon limit of the extremal  equally rotating Myers-Perry black hole in even dimensions \cite{Tanahashi:2012si}.

One can see that the critical rotation becomes larger than the upper bound on the rotation of Anti de Sitter solutions
%
\begin{eqnarray}
a_{c}^{2} >-\frac{1}{\hat{\Lambda}}
\end{eqnarray}
%
for all $\ell$ at $\hL<-2$. The equally rotating Myers-Perry black hole in Anti de Sitter should satisfy $a^{2}<-\hL^{-1}$ as mentioned in eq. (\ref{aLB}). So the large Anti de Sitter black hole with $\hL<-2$ does not show the dynamical instabilities. 
%
\begin{figure}[t]
 \begin{center}
 \includegraphics[width=70mm,angle=0]{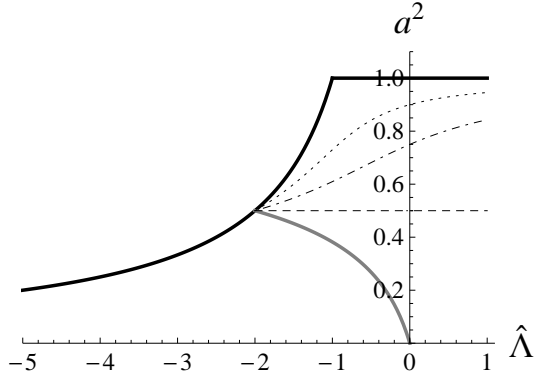}
 \end{center}
 \vspace{-5mm}
 \caption{ The available region of solutions and unstable region are shown for even dimensional solution. The black thick line shows the upper bound on the rotation for regular subextremal solution. The dashed, dotdashed and dotted line show the critical rotation for $\ell=2$, $\ell=4$ and $\ell=10$ respectively. The region between the upper bound and the critical rotation is the unstable region. The gray line shows the superradiant curve $\Omega_{\tH}(-\hL)^{-1/2}=1$ for Anti de Sitter solution . In the region above the gray line the solution can be superradiantly unstable.
}
 \label{fig:aL}
\end{figure}
%
This can be seen in Figure \ref{fig:aL}. In Figure \ref{fig:aL} we plot the rigid upper bound on the rotation and extremal limit by the black line. The dashed, dotdashed and dotted line show the critical rotation for $\ell=2$, $\ell=4$ and $\ell=10$ mode respectively. The region between the upper bound line and critical rotation line is the unstable region. The gray line is the superradiant curve \cite{Kunduri:2006qa} defined by
%
\begin{eqnarray}
\Omega_{\tH}(-\hL)^{-1/2}=1,
\end{eqnarray}
%
for Anti de Sitter solution. In the region above the gray line, the Myers-Perry black hole in Anti de Sitter can be superradiantly unstable.  We can see that the large AdS black hole does not have unstable region in even dimensions, and it is dynamically stable. This is similar feature with singly rotating Myers-Perry black hole in Anti de Sitter \cite{Dias:2010gk}\footnote{The difference is that the singly rotating Myers-Perry black hole does not have the extremal limit. }. The even dimensional equally rotating Myers-Perry black hole in de Sitter or flat spacetime is always dynamically unstable at larger rotation than the critical rotation.




\end{document}